\documentclass[manuscript,preprint2]{aastex}
\def\gs{\mathrel{\raise0.35ex\hbox{$\scriptstyle >$}\kern-0.6em \lower0.40ex\hbox{{$\scriptstyle \sim$}}}}
\def\ls{\mathrel{\raise0.35ex\hbox{$\scriptstyle <$}\kern-0.6em \lower0.40ex\hbox{{$\scriptstyle \sim$}}}}
\newcommand{\Msolar}{\mbox{$\rm M_{\odot}\,$}}
\newcommand{\Lsolar}{\mbox{$\rm L_{\odot}\,$}}
              
\newcommand{\arcsecs}{\mbox{$^{\prime\prime}$}}

\shorttitle{Submm properties of near-IR selected galaxies}
\shortauthors{Greve et al.}

\begin{document}


\title{A LABOCA survey of the Extended Chandra Deep Field South -- 
submillimeter properties of near-infrared selected galaxies}


\author{T.R.\ Greve\altaffilmark{1}, 
A.\ Wei\ss\altaffilmark{2}, 
F.\ Walter\altaffilmark{1}, 
I.\ Smail\altaffilmark{3}, 
X.Z.\ Zheng\altaffilmark{4},
K.K.\ Knudsen\altaffilmark{5}, 
K.E.K.\ Coppin\altaffilmark{3},
A.\ Kov\'{a}cs\altaffilmark{2},
E.F.\ Bell\altaffilmark{1},
C.\ de Breuck\altaffilmark{6},
H.\ Dannerbauer\altaffilmark{1},
M.\ Dickinson\altaffilmark{7},
E.\ Gawiser\altaffilmark{8},
D.\ Lutz\altaffilmark{9},
H.-W.\ Rix\altaffilmark{1}, 
E.\ Schinnerer\altaffilmark{1},
D.\ Alexander\altaffilmark{3},
F.\ Bertoldi\altaffilmark{5},
W.N.\ Brandt\altaffilmark{10},
S.C.\ Chapman\altaffilmark{11},
R.J.\ Ivison\altaffilmark{12},
A.M.\ Koekemoer\altaffilmark{13},
E.\ Kreysa\altaffilmark{2},
P.\ Kurczynski\altaffilmark{8},
K.\ Menten\altaffilmark{2},
G.\ Siringo\altaffilmark{2},
M.\ Swinbank\altaffilmark{3},
P.\ van der Werf\altaffilmark{14}
}
\email{tgreve@mpia-hd.mpg.de}

\altaffiltext{1}{Max-Planck-Institut f\"{u}r Astronomie, 69117 Heidelberg, Germany}
\altaffiltext{2}{Max-Planck-Institut f\"{u}r Radioastronomie, D-53121 Bonn, Germany}
\altaffiltext{3}{Institute for Computational Cosmology, Durham University, Durham DH1\,6BH, UK}
\altaffiltext{4}{Purple Mountain Observatory, Chinese Academy of Sciences, Nanjing 210008, China}
\altaffiltext{5}{Argelander Institute for Astronomy, University of Bonn, 53121 Bonn, Germany}
\altaffiltext{6}{European Southern Observatory, Garching bei M\"{u}nchen, Germany}
\altaffiltext{7}{National Optical Astronomical Observatory, Tucson, Arizona 85719, USA}
\altaffiltext{8}{Physics \& Astronomy Department, Rutgers University, Piscataway, NJ 08854, USA} 
\altaffiltext{9}{Max-Planck-Institut f\"{u}r extraterrestrische Physik, 85741 Garching bei M\"{u}nchen, Germany}
\altaffiltext{10}{Pennsylvania State University, Astronomy and Astrophysics 525 Davey Lab, University Park, PA 16802, USA}
\altaffiltext{11}{Institute for Astronomy, Cambridge CB3 0HA, UK}
\altaffiltext{12}{UK Astronomy Technology Centre, Royal Observatory, Edinburgh EH9\,3HJ, UK}
\altaffiltext{13}{Space Telescope Science Institute, Baltimore, Maryland 21218, USA}
\altaffiltext{14}{Leiden Observatory, PO Box 9513, 2300 RA Leiden, The Netherlands}


\begin{abstract}
Using the 330\,hr ESO-MPG 870-$\mu$m survey of the Extended Chandra Deep Field South (ECDF-S)
obtained with the Large Apex BOlometer CAmera (LABOCA) on the Atacama Pathfinder EXperiment (APEX), 
we have carried out a stacking analysis at submillimeter (submm)
wavelengths of a sample of 8266 near-infra-red (near-IR) selected
($K_{\mbox{\tiny{vega}}}\le 20$) galaxies, including 893 BzK galaxies, 1253
extremely red objects (EROs) and 737 distant red galaxies (DRGs), selected from
the Multi-wavelength Survey by Yale-Chile (MUSYC).  We measure average
870-$\mu$m fluxes of $0.20\pm 0.01\,$mJy ($20.0\sigma$), $0.45\pm 0.04\,$mJy
($11.3\sigma$), $0.42\pm 0.03\,$mJy ($14.0\sigma$), and $0.41\pm 0.04\,$mJy
($10.3\sigma$) for the $K_{\mbox{\tiny{vega}}}\le 20$, BzK, ERO and DRG samples,
respectively. For the BzK, ERO and DRG subsamples, which overlap to some degree
and are like to be at $z\simeq 1-2$, this implies an average far-IR luminosity
of $\sim 2-6\times 10^{11}\,\Lsolar$ and star formation rate of $\sim 40-100\,\Msolar$.
Splitting the BzK galaxies up into star-forming (sBzK) and
passive (pBzK) galaxies, the former is significantly detected ($0.48\pm
0.04\,$mJy, $12.0\sigma$) while the latter is only marginally detected
($0.27\pm 0.10\,$mJy, $2.7\sigma$), thus confirming that the sBzK/pBzK criteria
do isolate obscured, star forming and truly passive galaxies.
The $K_{\mbox{\tiny{vega}}}\le 20$
galaxies are found to contribute with $6.6\pm 0.3\,$Jy$\,$deg$^{-2}$ ($\sim
15\,\%$) to the 870-$\mu$m extragalactic background light (EBL). sBzK and pBzK
galaxies contribute $1.7\pm 0.2\,$Jy$\,$deg$^{-2}$ ($\sim 4\,\%$) and $0.2\pm
0.1\,$Jy$\,$deg$^{-2}$ ($< 0.5\,\%$) to the EBL. We present the first
delineation of the average submm signal from $K_{\mbox{\tiny{vega}}}\le 20$
selected galaxies and their contribution to the submm EBL as a function of
(photometric) redshift, and find a decline in the average submm signal (and
therefore IR luminosity and star formation rate) by a factor $\sim 2-3$ from
$z\sim 2$ to $z\sim 0$. This is in line with a cosmic star formation history in
which the star formation activity in galaxies increases significantly at $z\gs 1$.  
A linear correlation between the average 24-$\mu$m and 870-$\mu$m flux densities
is found for $K_{\mbox{\tiny{vega}}}\le 20$ galaxies with 24-$\mu$m fluxes $\ls
350\,\mu$Jy (corresponding to $L_{\mbox{\tiny{IR}}}\simeq 1.5\times
10^{12}\,\Lsolar$ a $z\simeq 2$), while at higher 24-$\mu$m fluxes there
is no correlation. This behaviour suggests that star formation, and not
Active Galactic Nuclei (AGN), is in general responsible for the bulk of the
mid-IR emission of $L_{\mbox{\tiny{IR}}}\ls 1.5\times 10^{12}\,\Lsolar$
systems, while in more luminous systems the AGN makes a significant
contribution to the 24-$\mu$m emission.  By mapping the stacked 870-$\mu$m
signal across the $B-z$ vs.\ $z-K$ diagram we have confirmed the ability of the
sBzK-selection criterion to select starforming galaxies at $z > 1$, although
our analysis suggest that the subset of sBzK galaxies which are also EROs are
responsible for $ > 80\,\%$ of the submm emission from the entire sBzK
population.
\end{abstract}


\keywords{cosmology: observations --- galaxies: evolution --- galaxies: 
formation --- galaxies: high-redshift}



\section{Introduction}
Extragalactic blank-field submm surveys have been carried out since
the advent of SCUBA (Holland 1999) more than a decade ago, and have provided us
with a unique view of intense, dust-cloaked star formation events at high
redshifts (e.g.\ Blain et al.\ 2002). Yet such observations have so far only
pin-pointed the most luminous high-$z$ galaxies, due to the limitations in
sensitivity and resolution imposed by preset-day (sub)mm facilities.  It is now
well-established that the bright ($\gs 5$\,mJy at 850-$\mu$m) submm sources
uncovered by these surveys primarily reside in the redshift range $z\simeq 1.5-3.5$
(Chapman et al.\ 2003, 2005), and account for $\sim 20-30$ percent of the
extragalactic background light (EBL) at 850-$\mu$m (Barger et al.\ 1998, 1999;
Hughes et al.\ 1998; Coppin et al.\ 2006). Surveys that make use of galaxy
clusters' gravitational amplification of the background source
plane have uncovered a number of faint ($S_{850\mu m} \gs 2$\,mJy)
sources and resolved up to 80 per-cent of the background (Smail et al.\ 1997,
2002a; Blain et al.\ 1999; Cowie et al.\ 2002; Chapman et al.\ 2002; Knudsen et
al.\ 2008). However, we know little about the nature and redshift distribution (Smail et al.\
1997, 2002a) of the population below $\sim 5\,$mJy due in part to the difficulty of identifying
counterparts in the radio.

The recent advent of large format near-IR cameras have revealed populations of
moderately star forming galaxies at $z\simeq 1-3$ that are more numerous than
the (sub)mm selected systems, and more representative of the bulk population at
these epochs (e.g.\ Cimatti et al.\ 2002; Lawrence et al.\ 2007).  The
rest-frame near-IR is arguably the best wavelength range to undertake such
surveys at -- as, in comparison with UV and optical surveys, it is less
sensitive to the effects of age and dust on the stellar population, and thus
more closely provides a selection based on stellar mass.  

Well known examples of near-IR colour-selected galaxies are the populations of
extremely red objects (EROs -- Elston, Rieke \& Rieke 1988, McCarthy, Persson
\& West 1992; Hu \& Ridgway 1994), distant red galaxies (DRGs -- Franx et al.\
2003; van Dokkum et al.\ 2003), and the so-called BzK galaxies (Daddi et al.\
2004). While these populations are selected according to different
optical/near-IR colour criteria applied, which pick out systems at different,
but overlapping, redshift ranges.  These colour criteria are often designed to
straddle the 4000\,\AA~break (including the Balmer break at 3646\,\AA),
characteristic of evolved, metal-enriched galaxies that are old enough that
OB-stars do not dominate the light.  The same colour criteria, however, will
also select dusty, starforming galaxies at virtually any redshift. Thus,
near-IR colour-selected galaxy populations are typically a mix of actively star
forming galaxies and old, evolved systems, which means additional colour
criteria and/or spectral analysis has to be applied in order to separate the
two. Clearly, submm observations offer a unique way of distinguishing between
starforming and passive near-IR galaxies.

At present, however, the bulk of near-IR colour-selected galaxies are too faint for
individual detection by large format (sub)mm surveys, and for the moment,
therefore, the only way forward is to study their average (sub)mm/far-IR
properties by means of stacking techniques.  A handful of such studies have
been carried out to date, characterizing the average submm signal of near-IR
selected galaxies and their contribution to the extragalactic background light
(EBL) at submm wavelengths (Webb et al.\ 2004; Daddi et al.\ 2005; Knudsen et
al.\ 2005; Takagi et al.\ 2007).  Yet most of these stacking analyses have been
of relatively small samples of galaxies, and as a consequence have had to
averaging their submm properties over the entire redshift range from which they
are selected (which is often substantial, $z\sim 1-3$). A robust delineation of
the submm signal of near-IR selected galaxies as a function of redshifts has
therefore been lacking, and as a consequence we do not know how the
dust-enshrouded starformation in these galaxies evolve with cosmic epoch.

The Extended Chandra Deep Field South (ECDF-S), a $\rm 30\arcmin \times
30\arcmin$ region centered on the smaller GOODS-S/CDF-S field (Giavalisco et
al.\ 2004), is one of the most intensively studied extragalactic fields in the
southern sky. In addition to X-ray observations with Chandra (Alexander et al.\
2003; Lehmer et al.\ 2005; Luo et al.\ 2008), the EDCF-S has been targeted in a
large number of optical and near-IR filter passbands from the ground as part of
COMBO-17 (Wolf et al.\ 2001) and MUSYC (Gawiser et al.\ 2003), and with HST/ACS
as part of GEMS (Rix et al.\ 2004).  Furthermore, deep mid-IR imaging has been
provided by the {\it Spitzer} IRAC/MUSYC Public Legacy in ECDF-S (SIMPLE, Damen et al.\ 2009), and
the {\it Spitzer}/MIPS Far-Infrared Deep Extragalactic Legacy Survey (FIDEL --
Dickinson et al.\ in prep. See also {\tt http://ssc.spitzer.caltech.edu/legacy/abs\\
/dickinson2.html}).

To study the submm properties of the sources in the ECDF-S, we have undertaken
a large ESO-MPG survey (Coppin et al.\ 2009; Wei\ss~et al.\ 2009) using the LABOCA 870-$\mu$m 
camera (Siringo et al.\ 2009) mounted on the Atacama Pathfinder Experiment (APEX) and combined the data
with the already existing multi-wavelength data available for this field.

Throughout this paper we adopt a flat cosmology with $\rm \Omega_M = 0.27,
\Omega_{\Lambda} = 0.73$, and $\rm h = 0.71$ (Spergel et al.\ 2003).

\section{The submm data}\label{section:submm-data}
Observations were carried out using the 295 horn-bolometer array LABOCA on APEX
(Siringo et al.\ (2009), and are discussed in detail in Wei\ss~et al.\ (2009)). 
The bolometers are AC-biased, operated in total power mode, and
distributed in a hexagonal configuration over the 11.4\arcmin~field of view.
The center frequency of LABOCA is 345\,GHz and its passband has a FWHM of $\rm
\sim60$\,GHz. The measured angular resolution is 19.2\arcsecs~(FWHM).  The
observations were carried out between May 2007 and November 2008 in mostly excellent
weather conditions (PWV typically 0.5 mm corresponding to a zenith opacity of
0.2 at the observing wavelength). The mapping pattern was chosen to give a
uniform coverage across a $30'\times30'$ area centered at Ra: $03^h32^m29^s$
Dec $-27^{\circ}48^{'}47^{''}$ (J2000). 

Mapping was performed alternating rectangular on-the-fly scans with raster of
spiral patterns.  For the latter mode the telescope traces in two scans spirals
with radii between $2'$ and $4'$ at 16 and 9 positions (the raster) spaced by
600\arcsecs~in azimuth and elevation. The scanning speed was typically between
2-3 arcmin per second. 

Calibration was achieved through observations of Mars, Uranus and Neptune as
well as secondary calibrators and was found to be accurate within 8.5\%. The
atmospheric attenuation was determined via skydips every $\sim$ 2 hours as well
as from independent data from the APEX radiometer which measures the line of
sight water vapor column every minute.  Focus settings were typically determined
once per night and checked during sunrise. Pointing was checked on nearby
quasars PMNJ0457-2324, PMNJ0106-4034 and PMNJ0403-3605 and found to be stable
within 3\arcsecs.

The data was reduced using the BoA reduction package (Schuller et al.\ in
prep.).  Individual maps were co-added (noise-weighted) and the final map was
beam-smoothed, resulting in a spatial resolution of 27\arcsecs~(FWHM).  The
total on-source observing time in the data used for this analysis is 200 hours
(330\,hr including overheads) and the average rms across the entire
$30'\times30'$ field is 1.2\,mJy beam$^{-1}$, making it the largest contiguous
(sub)mm survey ever undertaken to this depth (c.f.\ Coppin et al.\ 2006;
Bertoldi et al.\ 2007).

\section{Near-IR selected galaxies}\label{section:near-IR-samples} 
We use the Wide MUSYC public data release of $UBVRIzJHK$ catalogues in the
ECDF-S to construct samples of near-IR selected galaxies (Taylor et al.\
2008)\footnote{The catalogue is available at {\it
http://www.astro.yale.edu/MUSYC/}}. The MUSYC survey covers the central $\rm
30\arcmin \times 30\arcmin$ of the ECDF-S and reaches 5-$\sigma$ point source
sensitivities in the near-IR of $J_{\mbox{\tiny{AB}}} = 22.7$ and

\section{Near-IR selected galaxies}\label{section:near-IR-samples} 
We use the Wide MUSYC public data release of $UBVRIzJHK$ catalogues in the
ECDF-S to construct samples of near-IR selected galaxies (Taylor et al.\
2008)\footnote{The catalogue is available at {\it
http://www.astro.yale.edu/MUSYC/}}. The MUSYC survey covers the central $\rm
30\arcmin \times 30\arcmin$ of the ECDF-S and reaches 5-$\sigma$ point source
sensitivities in the near-IR of $J_{\mbox{\tiny{AB}}} = 22.7$ and
$K_{\mbox{\tiny{AB}}} = 22.0$, respectively (see Gawiser et al.\ (2006) for a
detailed description of the MUSYC survey). 

\smallskip

\begin{deluxetable}{l|llllll}
\tabletypesize{\scriptsize}
\tablecaption{{\small 
The total number of sources in each sample (given in parentheses in the first column of each line), followed
by the percentages contributed by the other samples. The latter are also given as absolute numbers in parentheses.}} 
\tablewidth{0pt}
\tablehead{
\colhead{  }                          & \colhead{sBzK} & \colhead{pBzK} & \colhead{ERO} & \colhead{DRG} & \colhead{sBzK+pBzK} & \colhead{ERO+DRG} 
}
\startdata
$K_{\mbox{\tiny{vega}}}\le 20$ (8266) &   9\%(744)     &   1.8\%(149)  & 15.2\%(1253)  &  8.9\%(737)         & 10.8\%(893)         & 19.6\%(1620)\\
sBzK                (744)             & 100\%(744)     &   0\%(0)       &  30.4\%(226)  & 36.2\%(269)   & 83.3\%(744)         & 49.7\%(370)\\
pBzK                (149)             &   0\%(0)       & 100\%(149)     &  98.0\%(146)  &  43.6\%(65)   & 16.7\%(149)         & 98\%(146)\\
ERO                (1253)             & 18.0\%(226)    & 11.7\%(146)    & 100\%(1253)   & 29.5\%(370)   & 29.7\%(372)         & 100\%(1253)\\
DRG                 (737)             & 36.5\%(269)    &  8.8\%(65)     &  50.2\%(370)  & 100\%(737)    & 45.3\%(334)         & 100\%(737)\\
sBzK+pBzK           (893)             & 83.3\%(744)    &  16.7\%(149)   &  41.7\%(372)  & 37.4\%(334)   & 100\%(893)          & 57.8\%(516)\\
ERO+DRG            (1620)             & 22.8\%(370)    &  9.0\%(146)    &  63.0\%(1253) & 37.0\%(737)   & 31.9\%(516)         & 100\%(1620)\\
\enddata
\label{table:overlap} 
\end{deluxetable}

\begin{figure*}[t]
\includegraphics[angle=0,scale=0.42]{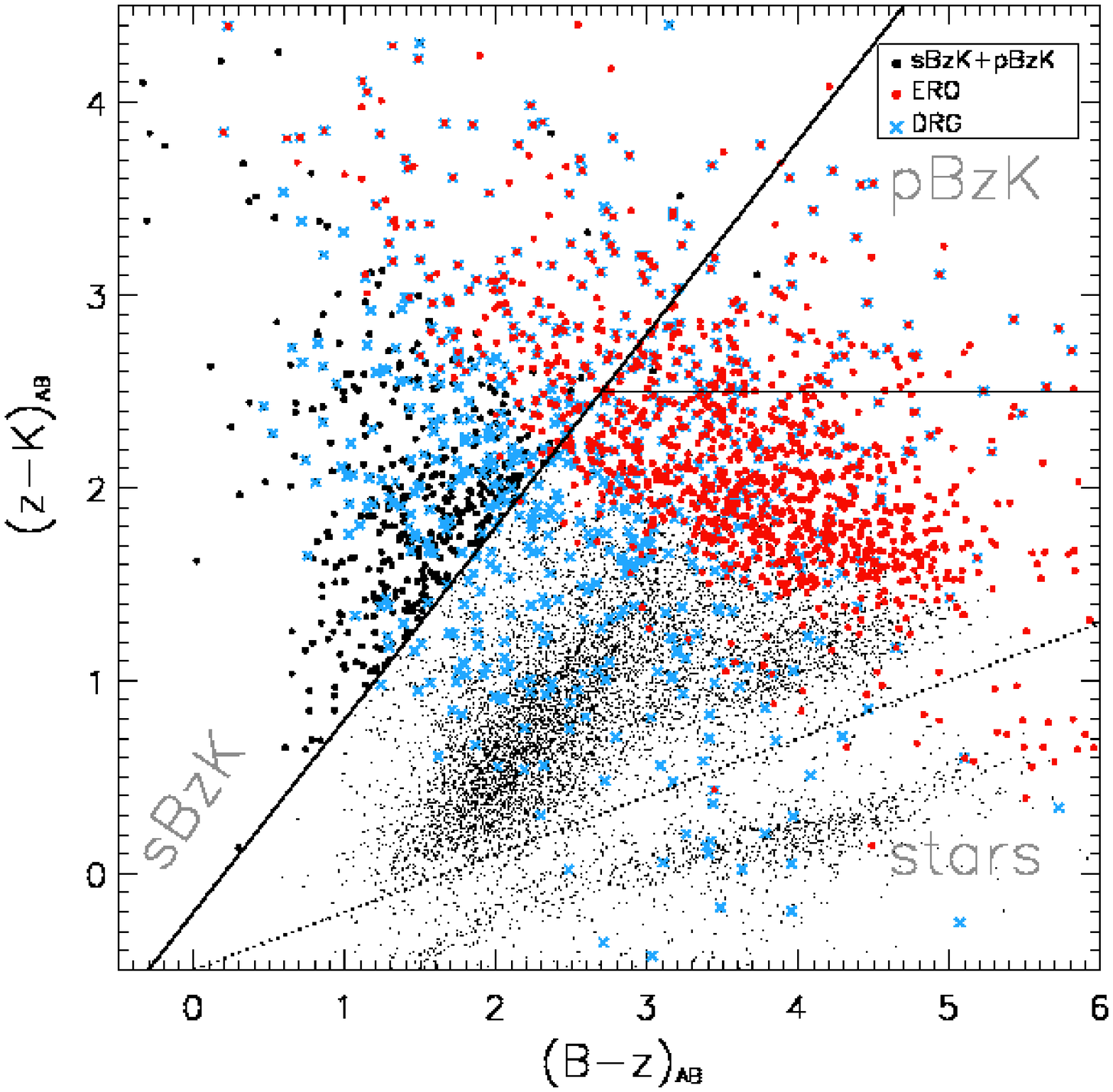}\includegraphics[angle=0,scale=0.42]{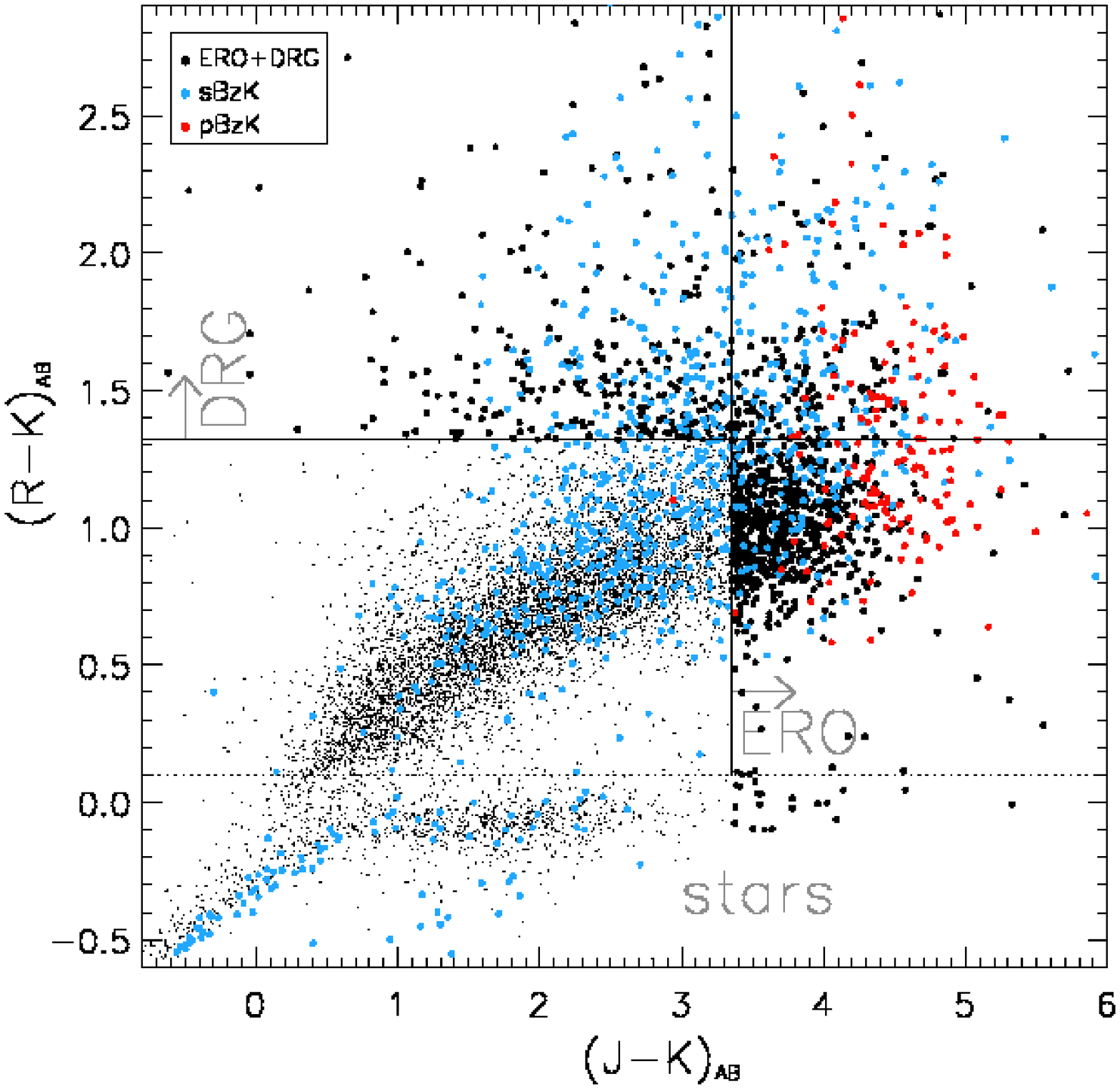}
\caption{Colour-magnitude diagrams illustrating the selection criteria for the
BzK, ERO and DRG samples.  {\bf Left:} The BzK-diagram (D04) showing the sBzK
and pBzK regions and the location of sources selected according to the ERO- and
DRG-criteria.  {\bf Right:} RJK diagram showing the ERO and DRG regions and the
location of the sources fulfilling the sBzK and pBzK colour criteria.  Sources
below the dotted lines in either diagram were deemed to be stars and therefore
discarded from the analysis. Note this includes a number of sources formally
selected as either BzK, ERO or DRG galaxies (see
\S~\ref{section:near-IR-samples}). The diagrams show the significant overlap
between the different near-IR selected populations.
} 
\label{figure:cc-diagram} 
\end{figure*}

We defined our sample as sources with $K_{\mbox{\tiny{vega}}}\le 20$
(corresponding to $K_{\mbox{\tiny{AB}}}\le 21.9$, i.e.\ we have assumed a 
VEGA-AB offset of 1.9 in the $K$-band). This magnitude cut-off was
chosen since the MUSYC survey is close to 100\% complete at this depth, and
since other studies have adopted the same magnitude limit, thus facilitating a
direct comparison.  Contamination by stars was accounted for by removing
objects lying on the stellar loci in the $(z-K)_{\mbox{\tiny{AB}}}$ vs.\
$(B-z)_{\mbox{\tiny{AB}}}$ or $(J-K)_{\mbox{\tiny{AB}}}$ vs.\
$(R-K)_{\mbox{\tiny{AB}}}$ diagrams (hereafter referred to as BzK and RJK
diagrams -- Fig.\ \ref{figure:cc-diagram}).  As shown in Fig.\
\ref{figure:cc-diagram}a and b, a non-negligible fraction of sources that
appear as normal galaxies in the BzK diagram end up in the stellar region in
the RJK diagram and vice versa. These sources were all poorly fit ($\chi^2/\nu_{\mbox{\tiny{dof}}} >
100$) by galaxy SED templates, thus strongly suggesting that they are in fact
stars and illustrating the need of removing both stellar loci from the sample.
This left us with a total sample of 8266 $K_{\mbox{\tiny{vega}}}\le 20$
selected galaxies.  From this sample, subsets of BzK galaxies, extremely red
objects (EROs) and distant red galaxies (DRGs) were extracted as described
below.

\smallskip

Applying the star-forming and passive BzK criteria, namely $BzK =
(z-K)_{\mbox{\tiny{AB}}} - (B-z)_{\mbox{\tiny{AB}}} \ge -0.2$ for sBzK
galaxies, and $BzK < -0.2$ and $(z-K)_{\mbox{\tiny{AB}}} > 2.5$ for pBzK
galaxies (Daddi et al.\ 2004), to our $K_{\mbox{\tiny{vega}}}\le 20$ catalogue
defined above, we obtained samples of 744 sBzK and 149 pBzK
galaxies\footnote{We matched the stellar sequence in the BzK-diagram with that
of Daddi et al.\ (2004) using $(z-K)_{D04} = (z'-K) - 0.2$.}. Sources which
were undetected at the 1$\,\sigma$ level in both $B$ and $z$ were not included.
Furthermore, sources formally belonging to the sBzK-region but which were
undetected in $B$, thus having $B-z$ lower limits only, were discarded, as were
sources in the pBzK region with lower limits in $z-K$.  A total of 1253 EROs
were selected from our $K_{\mbox{\tiny{vega}}}\le 20$ sample using the standard
criterion, i.e.\ $(R-K)_{\mbox{\tiny{AB}}} \ge 3.35$ (equivalent to
$(R-K)_{\mbox{\tiny{vega}}}\gs 5$ -- Elston, Rieke \& Rieke 1988; McCarthy,
Persson \& West 1992; Hu \& Ridgeway 1994).  Finally, the selection of DRGs was
done by applying the colour cut $(J-K)_{\mbox{\tiny{AB}}}>1.32$ (Franx et al.\
2003). In total 737 DRGs with $K_{\mbox{\tiny{vega}}}\le 20$ were selected in
this way.

\subsection{Overlap between BzK, ERO, and DRG samples}\label{subsection:overlap} 
In this paper we aim to determine the contribution to the submm background from
$K_{\mbox{\tiny{vega}}}\le 20$ galaxies, as well as the sub-samples of 
BzKs, EROs and DRGs and their joint contributions, and it is therefore
important to determine the degree of overlap between these populations. 
The overlaps in terms of percentages are given in Table
\ref{table:overlap}.

\begin{figure}[t]
\includegraphics[angle=0,width=1.0\hsize]{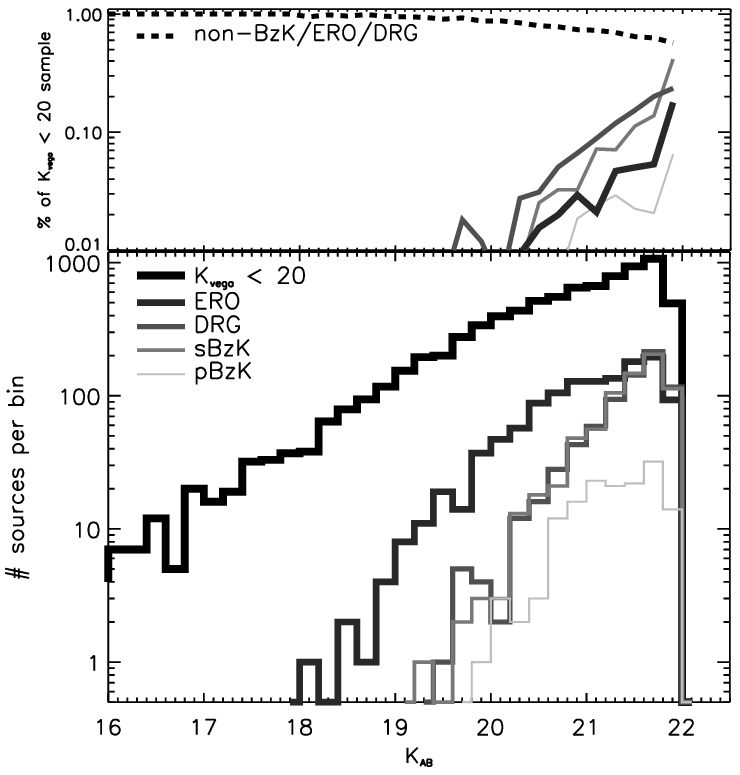}
\caption{{\bf Top:} Percentages, i.e.\ fraction of the full
$K_{\mbox{\tiny{vega}}}\le 20$ sample, of the BzK, ERO and DRG samples, as well
as the non-BzK/ERO/DRG sample (dashed line) as a function of their
$K_{\mbox{\tiny{AB}}}$ magnitude.  {\bf Bottom:} The number distribution of
sources as a function of their $K_{\mbox{\tiny{AB}}}$ magnitude. The
distributions for the full sample as well as the BzK, ERO and DRG samples are
shown in order to illustrate the overlap between the different populations.
While the BzK, ERO and DRG samples start making up a non-negligible fraction of
the parent sample for $K_{\mbox{\tiny{AB}}}\gs 20$, even at the faintest
magnitudes more than half of the sources from the parent sample do not fall
within any of these three classifications.
} 
\label{figure:overlap} 
\end{figure}

From Fig.\ \ref{figure:overlap} it is seen that BzK, ERO and DRG galaxies only
start to contribute significantly ($> 1\%$) to the full sample for
$K_{\mbox{\tiny{AB}}}\gs 20$. Even so, more than half of the full sample does
not fall within the BzK/ERO/DRG classifications at these faint flux levels.
Of the full $K_{\mbox{\tiny{vega}}}\le 20$ sample, 6269 sources (corresponding
to $\sim 76\%$) do not classify as BzKs, EROs or DRGs.

The locations of the various samples in the BzK and RJK diagrams are shown in
Fig.\ \ref{figure:cc-diagram}. The DRGs are seen to be spread out across the
BzK diagram, while the EROs lie in a much more well defined region of the BzK
diagram. EROs and DRGs make up about 30\% and 36\% of the sBzKs, respectively.
Similarly, we find that EROs and DRGs constitute 98\% and 44\% of the pBzK
sample, respectively.  Clearly, pBzK galaxies are much more often selected as
EROs than is the case for sBzKs, while the occurrence of a pBzK or sBzK galaxy
being classified as a DRG is about the same.  Turning to the RJK-diagram we see
that the sBzK galaxies are much more spread out than the pBzKs. About 18\% and
37\% of EROs and DRGs, respectively, are made up of sBzK, while pBzKs make up
about 12\% and 9\% of the two populations.

The overlaps between BzK, ERO and DRG galaxies have been discussed in detail in
other studies (e.g.\ Reddy et al.\ 2005; Grazian et al.\ 2007; Takagi et al.\
2007; Lane et al.\ 2008). The most statistically significant study was carried
out by Lane et al.\ (2009) who used the UKIRT Infrared Deep Sky Survey (UKIDSS)
Ultra Deep Survey Early Data Release (UDS EDR) to study large samples of BzK,
ERO and DRG galaxies. For samples selected down to $K_{\mbox{\tiny{AB}}}=21.2$, which is
close to our magnitude cut-off, they found sBzK:ERO and pBzK:ERO ratios of 32\%
and 95\%, respectively, i.e.\ in excellent agreement with our values. They also
find that about 30\% of DRGs are sBzK, again in good agreement with our findings (see
also Reddy et al.\ 2005).

\subsection{Redshift distributions}\label{subsection:phot-z} 
The sample was correlated against publically available
spectroscopic redshift surveys (Szokoly et al.\ 2004; Vanzella et al.\ 2005,
2006, 2008; Popesso et al.\ 2008; Kriek et al.\ 2008).  Using only the most
reliable spectroscopic redshifts from these surveys we extracted 2341 redshifts,
of which the majority lie within the CDF-S region. A total of 546 galaxies from
our sample were matched to a spectroscopic redshift. Of these 28 were
sBzK, 21 were ERO, and 10 were DRG galaxies.  The fact that no
pBzK galaxies were identified with a spectroscopic redshift is not too surprising
since these are in all likelihood old, evolved galaxies with optical spectra devoid of emission
features, thus making it difficult to obtain robust spectroscopic redshifts
(cf.\ Kriek et al.\ 2006; Cimatti et al.\ 2008).  Sources with spectroscopic
redshifts were used as a test sample to optimize input parameters for the
photometric redshift code EAZY (Brammer, van Dokkum \& Coppi 2008). The code
works by fitting non-negative linear combinations of galaxy spectra to the
observed spectral energy distributions (SEDs), which in our case consisted of the nine
MUSYC filter bands.  

\begin{figure}[t] 
\includegraphics[angle=0,width=1.0\hsize]{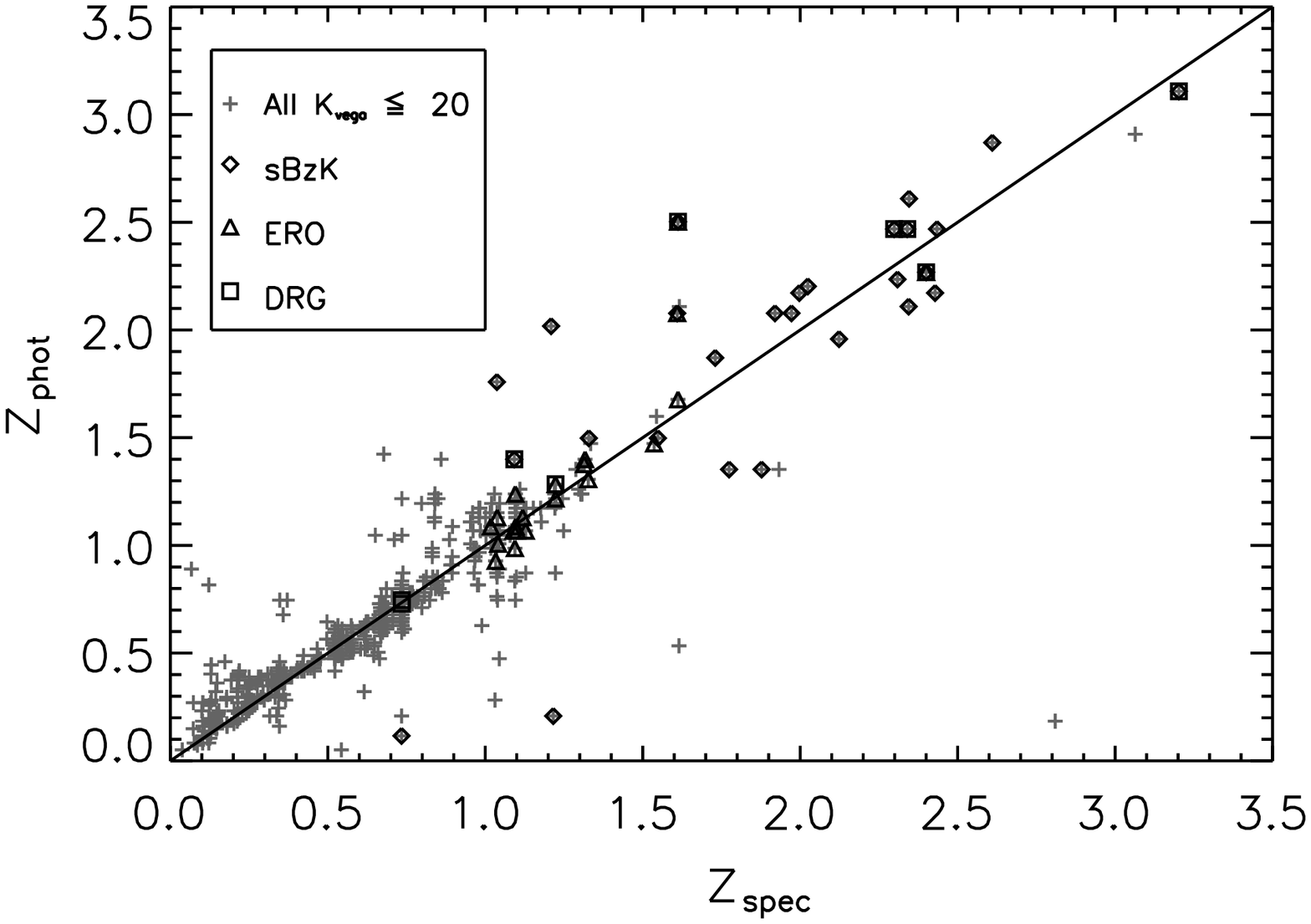} 
\caption{Photometric versus spectroscopic redshifts for the sources in the full $K_{\mbox{\tiny{vega}}}\le 20$ sample,
and the sBzK/pBzK, ERO and DRG sub-samples which have spectroscopic redshifts (see
\S~\ref{subsection:phot-z}).   
The normalized median absolute
deviation of $\Delta z = z_{\mbox{\tiny{phot}}}-z_{\mbox{\tiny{spec}}}$ (see
Brammer, van Dokkum \& Coppi 2008) is  $\simeq 0.037$ for $z\le 1.5$ and
$\simeq 0.079$ for $z> 1.5$.
} 
\label{figure:eazy} 
\end{figure}

The resulting photometric redshifts are compared against their spectroscopic
counterparts in Fig.\ \ref{figure:eazy}.  The normalized median absolute
deviation of $\Delta z = z_{\mbox{\tiny{phot}}}-z_{\mbox{\tiny{spec}}}$ (see
Brammer, van Dokkum \& Coppi 2008) is  $\simeq 0.037$ for $z\le 1.5$ and
$\simeq 0.079$ for $z> 1.5$.  Significant outliers, which we define to be
sources with $| \Delta z|/(1+z_{\mbox{\tiny{spec}}})$ 5 times greater than the
median, make up $\sim 9\%$ of the total sample.  These numbers are consistent
with the typical performances of photometric redshift codes (e.g.\ Bolzonella,
Miralles \& Pell\'{o} 2000; Quadri et al.\ 2007; Brammer van Dokkum \& Coppi
2008). 

\begin{figure}[t] 
\includegraphics[angle=0,width=0.83\hsize]{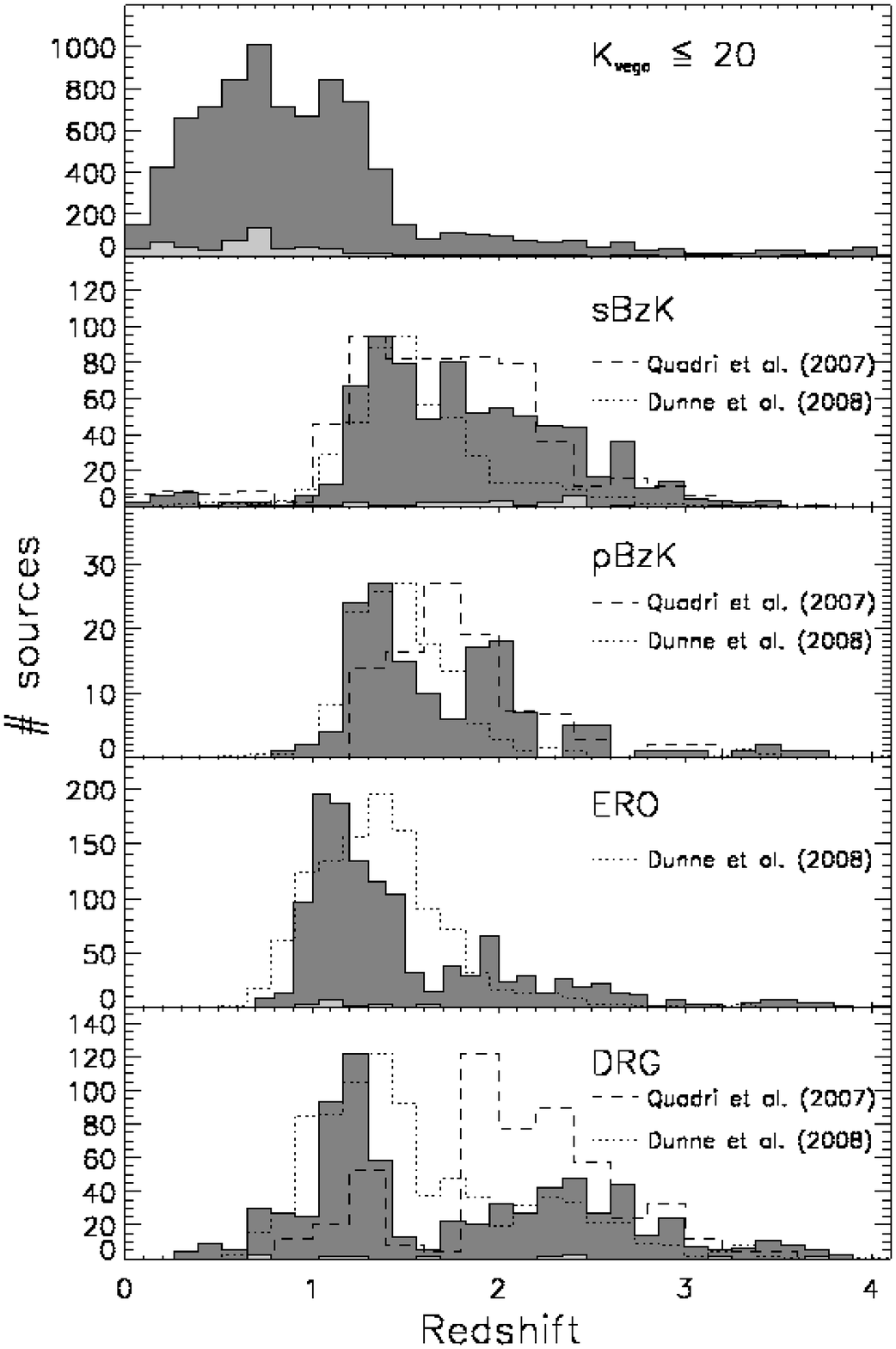} 
\caption{Top to bottom: redshift distributions of the
$K_{\mbox{\tiny{vega}}}\le 20$, sBzK, pBzK, ERO and DRG samples, respectively
(dark grey histograms). Spectroscopic redshifts are shown as light grey
histograms.  Also shown are the photometric redshift distributions of the
$K_{\mbox{\tiny{vega}}}\le 20$ sBzK, pBzK and DRG samples by Q07, as well as of
the $K_{\mbox{\tiny{vega}}}\le 20$ selected samples by D08.  The Q07 and D08
distributions have been scaled to match the peaks of our distributions.  Given
the uncertainties, our $K_{\mbox{\tiny{vega}}}\le 20$ BzK, ERO and DRG redshift
distributions are in good agreement with those of Q07 and D08.
}
\label{figure:Nz} 
\end{figure}

Adopting the parameters for the test sample, photometric redshifts were derived
for the remainder of the sample without spectroscopic redshifts.  The redshift
distributions of the full $K_{\mbox{\tiny{vega}}}\le 20$ sample as well as the
BzK, ERO, and DRG samples are shown in Fig.\ \ref{figure:Nz}, where we also
compare with two other photometric redshift distributions obtained from: 1) the
MUSYC UBVRI and deep JHK imaging of the three $10'\times 10'$ fields HDFS1/S2,
MUSYC 1030 and 1255 (Quadri et al.\ 2007 - hereafter Q07), and 2) deep
BVR$i'z'$JK imaging of 1113\,arcmin$^2$ in the Ultra-Deep Survey (UDS) portion
of the United Kingdom Infrared Telescope Deep Sky Survey (UKIDSS) (Dunne et
al.\ 2008 - hereafter D08).

The BzK selection criteria are defined to select galaxies in the redshifts
range $1.4 < z < 2.5$ (Daddi et al.\ 2004),  yet both the sBzK and pBzK
redshift distributions extend below and above this range. Of the sBzK galaxies,
66\,\% lie in the range $1.4< z < 2.5$, while 22\,\% are at $z< 1.4$ and 12\,\%
at $z> 2.5$. For the pBzK galaxies, the corresponding percentages are 60\,\%,
31\,\%, and 9\,\%, respectively. Both Q07 and D08 find similar fractions for
their samples of $K_{\mbox{\tiny{vega}}}\le 20$ galaxies, suggesting that the
BzK criteria select galaxies across a somewhat broader redshift range ($1\ls z
\ls 3.5$). 
The redshift distribution of EROs is seen to peak strongly at $z\simeq 1.1$
with a tail extending to $z\sim 3.5$. This is in line with photometric and
spectroscopic surveys which have shown that the redshift distribution of EROs
peaks around $z\simeq 1.2$ (Cimatti et al.\ 2003; Yan et al.\ 2004). The
redshift distribution of $K_{\mbox{\tiny{vega}}}\le 20$ EROs derived by D08
peaks at slightly higher redshifts ($z\sim 1.4$), but overall appears similar
to ours.  The DRG distribution shows prominent peaks at $z\simeq 1.2$ and
$z\simeq 2$, with the former being the most dominant. A similar bimodality is
also apparent in the $K_{\mbox{\tiny{vega}}}\le 20$ DRG sample by Q07, although
the dominant peak in their distribution lies at $z\simeq 2$.  The distribution
by D08 broadly resembles ours, with a prominent peak at $z\simeq 1.2$ followed
by a significant high-$z$ tail.  Overall, therefore, our redshift distributions
are consistent with those of Q07 and D08, given the uncertainties associated
with photometric redshift derivation, and the effects of field-to-field
variations.

\section{870-$\mu$m Stacking}\label{section:870um-stacking} 
In order to estimate the average 870-$\mu$m fluxes for the above excised
catalogues of near-IR galaxies we stack the 870-$\mu$m flux values
at their near-IR positions in the LABOCA map. 

\subsection{Submm-bright near-IR selected galaxies}\label{subsection:submm-bright-sources}
First, however, we need to identify any galaxies which are associated with
robust LABOCA sources, which we take to mean sources detected at $\ge
3.7$-$\sigma$ significance (see Wei\ss~et al.\ 2009 for details).  To this end
we adopt a search radius of 12.8\arcsecs around each LABOCA source, which
corresponds to the 95 per-cent confidence search radius given the
FWHM=19\arcsecs~beam. For a galaxy to qualify as a near-IR counterpart to a
LABOCA source, we furthermore require that $z > 0.8$. If more than one galaxy
meets these criteria for a given LABOCA source, we adopt the one closest to the
submm source.  In this manner we find 54 submm-near-IR associations from the
$K_{\mbox{\tiny{vega}}}\le 20$ sample.  Of these, 17/2 sources are classified
as sBzK/pBzK galaxies and 28 as EROs (of which 16 are also DRGs). 11 of the 17
sBzK galaxies are also EROs and of those, 9 are DRGs. Both of pBzK galaxies are
EROs, and also DRGs. As a safeguard against contamination, these submm-bright
near-IR sources were removed from the stacking analysis, although their
contribution was included (in a variance-weighted fashion) in the final tally
of average submm flux (Table \ref{table:stacking}).

\subsection{Stacking and deblending technique}\label{subsection:submm-stacking}
Next, we proceeded to perform a stacking analysis of the remaining submm-undetected
near-IR selected galaxies. Due to the slightly varying noise across the map, the average
870-$\mu$m flux ($\langle S_{\mbox{\tiny{870$\mu$m}}}\rangle$) and noise ($\langle
\sigma_{\mbox{\tiny{870$\mu$m}}}\rangle$) values were calculated as the variance-weighted mean,
i.e.\ 
\begin{equation} 
\langle S_{\mbox{\tiny{870$\mu$m}}}\rangle = \frac{\sum_i S_i/\sigma_i^2}{\sum_i 1/\sigma_i^2}, 
\label{equation:signal} 
\end{equation} 
and
\begin{equation} 
\langle \sigma_{\mbox{\tiny{870$\mu$m}}}\rangle = \frac{1}{\sqrt{\sum_i 1/\sigma_i^2}}, 
\label{equation:sigma} 
\end{equation} 
where $S_i$ and $\sigma_i$ are the 870-$\mu$m flux and r.m.s.\ noise pixel
values at the near-IR position of the $i^{th}$ source in the stack,
respectively.  To avoid the stack being contaminated by robust 870-$\mu$m
sources,  the stacking was performed on a 'residual' version of the LABOCA map
in which all of the 55 870-$\mu$m sources uncovered (Wei\ss~et al.\ 2009) had
been subtracted using a scaled beam profile. 

\begin{figure}[h] 
\includegraphics[angle=0,scale=0.42]{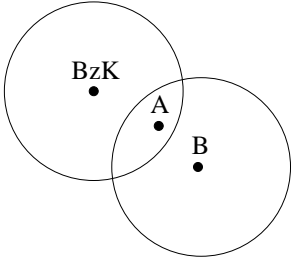} 
\caption{An example of where deblending is necessary. In order to properly
calculate the 870-$\mu$m flux coming from the BzK galaxy at its position we have
to calculate A's contribution, which in turn is affected by B. In total,
therefore, a linear set of three equations has to be solved (see eqs.\
3-5). The large circles indicate the LABOCA beam.}
\label{figure:deblend} 
\end{figure}

An important aspect of any stacking analysis performed on maps with coarse
angular resolution, is the issue of deblending of sources that lie within a
single resolution element.  For example, if a BzK galaxy has a neighbour, A,
within a LABOCA beam, we have to calculate the 870-$\mu$m flux contribution
from A at the position of the BzK galaxy. Of the full
$K_{\mbox{\tiny{vega}}}\le 20$ sample, we find this to be the case for 5985
sources (i.e.\ 72 per-cent of the sample). We correct for the blending of
sources by assuming Gaussian sources with FWHMs equal to the LABOCA beam. This
method is similar to the one adopted by Webb et al.\ (2004), although they (and
subsequent submm stacking studies) did not take into account the effects from
neighbour's neighbours.  An illustrative example of the latter is given Fig.\
\ref{figure:deblend} where A itself has a neighbour, B, within a LABOCA beam,
that is not within the LABOCA beam as measured from the BzK galaxy's position.
In this case we have to calculate B's contribution to A, in order to correctly
calculate A's contribution to the BzK galaxy, and the system of equations to be
solved is therefore 
\begin{eqnarray}
f_{\mbox{\tiny{BzK}}} &=& I_{\mbox{\tiny{BzK}}} + I_{\mbox{\tiny{A}}} e^{-r_{\mbox{\tiny{BzK,A}}}^2/(2\sigma^2)}\\ 
\label{equation:deblend-1}
f_{\mbox{\tiny{A}}}   &=& I_{\mbox{\tiny{A}}} + I_{\mbox{\tiny{BzK}}} e^{-r_{\mbox{\tiny{BzK,A}}}^2/(2\sigma^2)} +I_{\mbox{\tiny{B}}} e^{-r_{\mbox{\tiny{B,A}}}^2/(2\sigma^2)}\\ 
\label{equation:deblend-2}
f_{\mbox{\tiny{B}}}   &=& I_{\mbox{\tiny{B}}} + I_{\mbox{\tiny{A}}} e^{-r_{\mbox{\tiny{B,A}}}^2/(2\sigma^2)}
\label{equation:deblend-3}
\end{eqnarray} 
where $I$ and $f$ are the measured and deblended fluxes at the
relevant positions, respectively, and $r$ are the distances between the sources.
In order to estimate the error one makes by only deblending the fluxes from
neighbours within a beam, not taking into account neighbours' neighbours, we ran
the stacking analysis under both scenarios. We find that the deblending scheme
by Webb et al.\ (2004) can overestimate the fluxes by $\rm \sim 10\%$ compared to the
scheme described in this paper, but more typically the error is at the $\sim 5\,\%$ level.  
We emphasize that one has to not only deblend
neighbours within a certain galaxy population, but also across populations. 
Therefore, the deblending analysis was carried out on the full $K_{\mbox{\tiny{vega}}}\le 20$
sample. In doing so we have ignored sources with
$K_{\mbox{\tiny{vega}}} > 20$, and they have therefore not been included in the deblending analysis.
In the following section, however, we show that these fainter sources do indeed make a contribution
to the submm signal, which must be subtracted from the stacked fluxes. 

\subsection{Stacking the full samples}
\begin{figure*}[t] 
\includegraphics[angle=0,width=1\hsize]{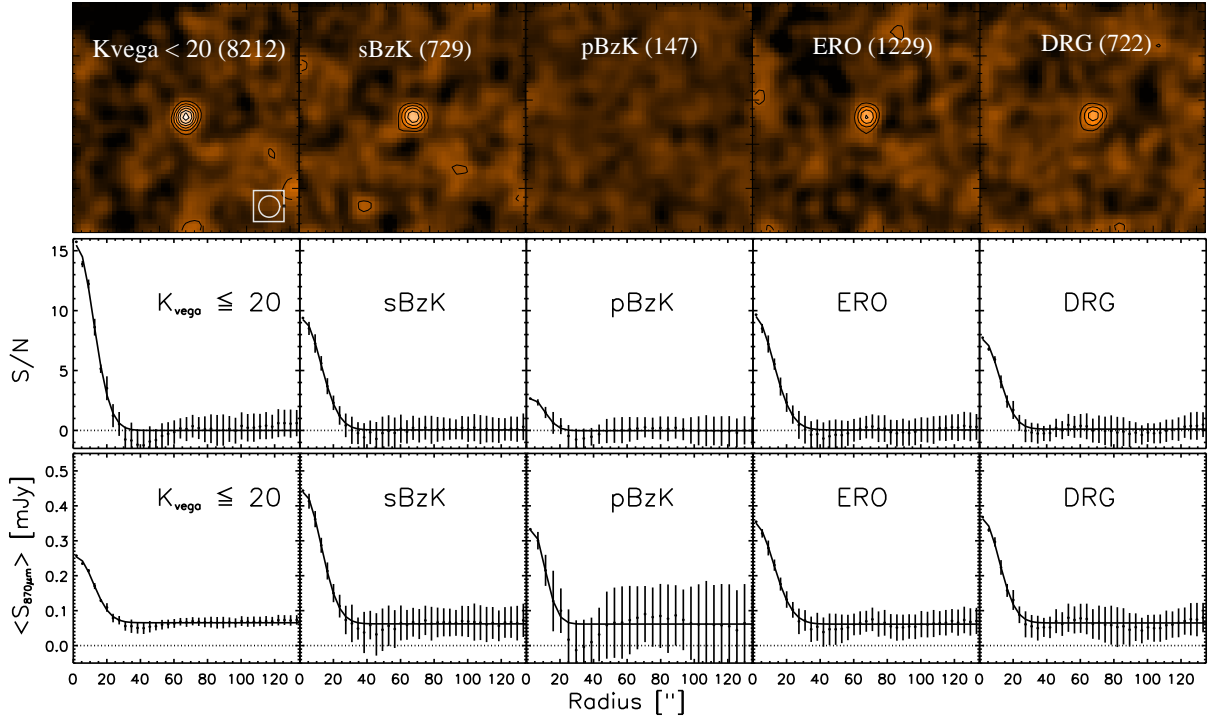} 
\caption{{\bf Top row:} Postage stamps of the stacked signal-to-noise images, each
$310\arcsecs \times 310\arcsecs$ in size, of the $K_{\mbox{\tiny{vega}}}\le
20$, BzK, ERO and DRG samples. The contours start at $S/N = 3$ and increase in
steps of 2. The number of sources in each stack is given in parentheses. The
angular resolution (FWHM=27\arcsecs) is shown as an insert in the left hand
panel. {\bf Middle row:} Radial profiles (azimuthally averaged) of the
corresponding stacked $S/N$ images (filled symbols). Gaussian fits to the profiles are 
indicated by the solid curve. {\bf Bottom row:} Radial profiles (azimuthally averaged) of the
corresponding stacked signal images (filled symbols). Gaussian fits to the profiles are 
indicated by the solid curve. Notice the non-zero baseline level ($0.065\,$mJy) caused
by the the $K_{\mbox{\tiny{vega}}} > 20$ sources with submm fluxes below the detection threshold.
This constant baseline level has been removed from the $S/N$ images and profiles in the top and
middle panels.
}
\label{figure:stack-images} 
\end{figure*}

For each source in the $K_{\mbox{\tiny{vega}}}\le 20$ sample the (deblended)
signal and noise values at its pixel position in the LABOCA map were recorded,
and from those the stacked 870-$\mu$m flux density of the full sample was
determined according to eqs.\ \ref{equation:signal} - \ref{equation:sigma}.
The 870-$\mu$m signal and noise values corresponding to the BzK, ERO and DRG
samples were extracted and their stacked 870-$\mu$m flux densities were derived
in a similar manner.
In a similar way, postage stamp images around each source were extracted and
combined in a weighted fashion resulting in stacked submm-images of the
$K$-selected samples (Fig.\ \ref{figure:stack-images}). From the
azimuthally-averaged radial profiles of the submm signal, it is clear that the
baseline level is not zero, but in fact there is a residual signal amounting to
$0.065\,$mJy and stemming from the population of $K_{\mbox{\tiny{vega}}} > 20$
galaxies lying below the submm detection limit of the LABOCA map.  In order to
account for this effect, the final stacked 870-$\mu$m flux densities (Table
\ref{table:stacking}) had a constant signal of $0.065\,$mJy subtracted from
them. As a comparison we also derived the median flux densities from the stacks
and found agreement (to within 15 percent) with the weighted averages.

\begin{deluxetable}{l|llll}
\tabletypesize{\scriptsize}
\tablecaption{{\small The stacked 870-$\mu$m flux densities for the $K_{\mbox{\tiny{vega}}}\le 20$ sample as well as the BzK, 
ERO, and DRG samples. The numbers of sources going into each stack are 
given in parentheses along with the significance of the stacked signals. Surface densities and contribution to
the 870-$\mu$m EBL are also listed.}} 
\tablewidth{0pt}
\tablehead{
\colhead{Galaxy type} & \colhead{$\rm \langle S_{870\mu m}\rangle^a$} & \colhead{$\rm \langle S_{870\mu m}\rangle^b$} & \colhead{$\rm \delta N^b$} & \colhead{$\rm \langle \delta I_{870\mu m}\rangle^b$}   \\ 
                      & ~~~~~~~~~~~~[mJy]                      & ~~~~~~~~~~~~[mJy]                                       & [sq.\ arcmin$^{-1}$]       & ~~~~~~~[Jy sq.\ deg$^{-1}$]                                   \\ 
}
\startdata
$K_{\mbox{\tiny{vega}}}\le 20$    & $\rm 0.19\pm 0.01$ (8212, $19.0\sigma$)  & $0.20\pm 0.01$ (8266, $20.0\sigma$)   & $9.18\pm 0.10$  &  $6.61\pm 0.34$ ($15.0\pm 5.2$\,\%)\\ 
sBzK                              & $\rm 0.38\pm 0.04$ (727, $9.5\sigma$)    & $0.48\pm 0.04$ (744, $12.0\sigma$)    & $0.83\pm 0.03$  &  $1.70\pm 0.24$ ($3.9\pm 1.4$\,\%)\\ 
pBzK                              & $\rm 0.21\pm 0.10$ (147, $2.1\sigma$)    & $0.27\pm 0.10$ (149, $2.7\sigma$)     & $0.17\pm 0.01$  &  $0.16\pm 0.11$ ($0.4\pm 0.3$\,\%)\\ 
sBzK+pBzK                         & $\rm 0.35\pm 0.04$ (874, $8.8\sigma$)    & $0.45\pm 0.04$ (893, $11.3\sigma$)    & $0.99\pm 0.03$  &  $1.89\pm 0.24$ ($4.3\pm 1.6$\,\%)\\ 
ERO                               & $\rm 0.30\pm 0.03$ (1225, $10.0\sigma$)  & $0.42\pm 0.03$ (1253, $14.0\sigma$)   & $1.39\pm 0.04$  &  $1.85\pm 0.20$ ($4.2\pm 1.5$\,\%)\\ 
DRG                               & $\rm 0.32\pm 0.04$ (717, $8.0\sigma$)    & $0.41\pm 0.04$ (737, $10.3\sigma$)    & $0.82\pm 0.03$  &  $1.09\pm 0.18$ ($2.5\pm 0.9$\,\%)\\ 
BzK/ERO/DRG$^c$                   & $\rm 0.31\pm 0.03$ (1961, $10.3\sigma$)  & $0.43\pm 0.03$ (1997, $14.3\sigma$)   & $2.22\pm 0.05$  &  $3.67\pm 0.29$ ($8.4\pm 2.9$\,\%)\\ 
non-BzK/ERO/DRG$^d$               & $\rm 0.16\pm 0.01$ (6251, $16.0\sigma$)  & $0.17\pm 0.01$ (6269, $17.0\sigma$)   & $6.97\pm 0.09$  &  $4.26\pm 0.26$ ($9.7\pm 3.3$\,\%)\\ 
\hline
                                              & & ~~~~~~~~~~~$z < 1.4$\\
\hline
$K_{\mbox{\tiny{vega}}}\le 20$                & $\rm 0.17\pm 0.01$ (7026, $17.0\sigma$)  & $0.17\pm 0.01$ (7055, $17.0\sigma$)   & $7.83\pm 0.09$  &  $5.36\pm 0.29$ ($11.7\pm 4.1\,$\%)\\ 
sBzK                                          & $\rm 0.33\pm 0.09$ (163, $3.7\sigma$)    & $0.42\pm 0.09$ (164, $4.7\sigma$)     & $0.18\pm 0.01$  &  $0.23\pm 0.14$ ($0.5\pm 0.4\,$\%)\\ 
pBzK                                          & $\rm -0.03\pm 0.17$ (47, $-0.18\sigma$)  & $-0.14\pm 0.17$ (47, $-0.8\sigma$)    & $0.05\pm 0.01$  &  ~~~~~~~~~~~~~$. . .$        \\ 
ERO                                           & $\rm 0.24\pm 0.04$ (740, $6.0\sigma$)    & $0.26\pm 0.04$ (751, $6.5\sigma$)     & $0.83\pm 0.03$  &  $0.87\pm 0.16$ ($1.9\pm 0.7\,$\%)\\ 
DRG                                           & $\rm 0.26\pm 0.06$ (351, $4.3\sigma$)    & $0.29\pm 0.06$ (373, $4.8\sigma$)     & $0.40\pm 0.02$  &  $0.47\pm 0.15$ ($1.0\pm 0.4$\,\%)\\ 
\hline
                                              & & ~~~~~~~~~~~$z > 1.4$\\
\hline
$K_{\mbox{\tiny{vega}}}\le 20$                & $\rm 0.34\pm 0.03$ (1186, $11.3\sigma$)  & $0.39\pm 0.03$ (1211, $13.0\sigma$)   & $1.35\pm 0.04$  &  $1.89\pm 0.20$ ($4.1\pm 1.5\,$\%)\\ 
sBzK                                          & $\rm 0.40\pm 0.05$ (564, $8.0\sigma$)    & $0.55\pm 0.05$ (580, $11.0\sigma$)    & $0.64\pm 0.03$  &  $1.28\pm 0.23$ ($2.8\pm 1.1\,$\%)\\ 
pBzK                                          & $\rm 0.32\pm 0.12$ (100, $2.7\sigma$)    & $0.41\pm 0.12$ (102, $3.4\sigma$)     & $0.11\pm 0.01$  &  $0.17\pm 0.16$ ($0.4\pm 0.4$\,\%)\\ 
ERO                                           & $\rm 0.40\pm 0.05$ (485, $8.0\sigma$)    & $0.54\pm 0.05$ (502, $10.8\sigma$)    & $0.56\pm 0.02$  &  $1.08\pm 0.22$ ($2.4\pm 0.1$\,\%)\\ 
DRG                                           & $\rm 0.39\pm 0.06$ (366, $6.5\sigma$)    & $0.59\pm 0.06$ (380, $9.8\sigma$)     & $0.42\pm 0.02$  &  $0.90\pm 0.23$ ($1.9\pm 0.8$\,\%)\\ 
\hline
                                              & & 24-$\mu$m detected ($S_{\mbox{\tiny{24$\mu$m}}} > 27\,\mu$Jy)\\
\hline
sBzK                                          & $\rm 0.55\pm 0.05$ (451, $11.0\sigma$)   & $0.70\pm 0.05$ (466, $14.0\sigma$)    & $0.52\pm 0.02$  &  $1.30\pm 0.27$ ($2.8\pm 1.1\,$\%)\\ 
pBzK                                          & $\rm 0.44\pm 0.16$ (52, $2.8\sigma$)     & $0.54\pm 0.16$ (53, $3.4\sigma$)      & $0.06\pm 0.01$  &  $0.11\pm 0.20$ ($0.2\pm 0.4\,$\%)\\ 
ERO                                           & $\rm 0.54\pm 0.05$ (490, $10.8\sigma$)   & $0.68\pm 0.05$ (511, $11.6\sigma$)    & $0.57\pm 0.03$  &  $1.39\pm 0.28$ ($3.0\pm 1.0\,$\%)\\ 
DRG                                           & $\rm 0.52\pm 0.06$ (379, $8.7\sigma$)    & $0.73\pm 0.06$ (400, $12.2\sigma$)    & $0.44\pm 0.02$  &  $1.16\pm 0.28$ ($2.5\pm 1.0\,$\%)\\ 
\hline
                                              & & 24-$\mu$m faint ($S_{\mbox{\tiny{24$\mu$m}}} < 27\,\mu$Jy)\\
\hline
sBzK                                          & $\rm 0.11\pm 0.07$ (276, $1.6\sigma$)    & $0.14\pm 0.07$ (278, $2.0\sigma$)     & $0.31\pm 0.02$  &  $0.16\pm 0.09$ ($0.3\pm 0.2\,$\%)\\ 
pBzK                                          & $\rm 0.08\pm 0.12$ (95, $0.7\sigma$)     & $0.12\pm 0.12$ (96, $1.0\sigma$)      & $0.11\pm 0.01$  &  $0.05\pm 0.06$ ($0.1\pm 0.1\,$\%)\\ 
ERO                                           & $\rm 0.14\pm 0.04$ (735, $3.5\sigma$)    & $0.18\pm 0.04$ (742, $4.5\sigma$)     & $0.82\pm 0.03$  &  $0.53\pm 0.14$ ($1.2\pm 0.5\,$\%)\\ 
DRG                                           & $\rm 0.10\pm 0.06$ (338, $1.7\sigma$)    & $0.18\pm 0.06$ (344, $3.0\sigma$)     & $0.38\pm 0.02$  &  $0.25\pm 0.11$ ($0.5\pm 0.2\,$\%)\\ 
\enddata
\label{table:stacking} 
\tablecomments{$^a$ Submm-faint sources only. $^b$ Submm-bright sources included. $^c$ The combined BzK, ERO and DRG samples, where source overlap between the populations have been
accounted for.} 
\end{deluxetable}

To gauge the significance of our results, we ran a series of Monte Carlo
simulations in which stacking analyses were carried out on 1000 versions of the
$K_{\mbox{\tiny{vega}}}\le 20$ catalogue, each with randomized positions with respect to the
original catalogue. Each source was assigned a random position by (randomly)
choosing a radius ($\rm 60\arcsecs \le r \le 200\arcsecs$) and an angle from its
original position. By confining the new positions within a certain distance of
the original, we ensured that the noise properties were similar to those in the
original stacking analysis.  As expected the distributions of stacked signals
obtained from these simulations were Gaussians centered on zero. We found that
the measured 870-$\mu$m signals occurred in $\rm < 0.05\%$ of the simulation runs.
Roughly the same percentage is found if we restrict our Monte Carlo analysis to
the sBzK, ERO and DRG samples, and clearly testify to the significance of the measured signals.

\begin{figure}[h]
\hspace*{-0.5cm}
\includegraphics[angle=0,scale=0.42]{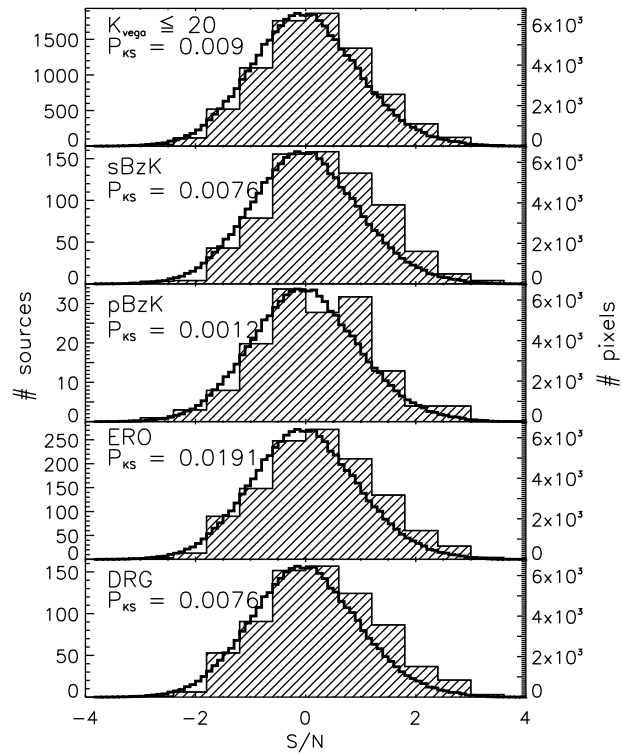} 
\caption{ The distributions of the $S/N$ values at the positions of the
$K_{\mbox{\tiny{vega}}}\le 20$ BzK, ERO, and DRG samples, shown as hashed
histograms from top to bottom, respectively.  The open histograms show the
$S/N$ distribution of the residual LABOCA map. The $P_{\mbox{\tiny{KS}}}$
values are the likelihoods that the sample distributions are identical to the
overall map distribution.}
\label{figure:snr-laboca-stack} 
\end{figure}

\smallskip

In addition to the above stacking analysis, we compared the distributions of
$S/N$ values at the near-IR source positions with the overall $S/N$
distribution of the residual LABOCA map (Fig.\ \ref{figure:snr-laboca-stack}).
In all cases, a formal Kolmogorov-Smirnov (KS) statistic shows that the sample
distributions are statistically different from the distribution of the entire
map which is symmetrical around zero (open histogram in Fig.\
\ref{figure:snr-laboca-stack}).  Unlike the distributions for the
$K_{\mbox{\tiny{vega}}}\le 20$, sBzK, ERO and DRG samples, which are clearly
biased towards positive $S/N$ values, the pBzK distribution shows outliers at
both positive and negative $S/N$ values. We can therefore not confidently claim
a detection even though this is formally suggest by the low KS probability.
This is in line with the above analysis which found that the stacked 870-$\mu$m
signal from pBzKs was not significant. 

\section{Discussion}\label{section:discussion}
\subsection{Stacked submm fluxes and star formation rates}\label{subsection:flux-SFR}
In \S~\ref{section:870um-stacking} we found that all of our $K$-selected
samples, except for the pBzKs, have significantly detected stacked 870-$\mu$m
fluxes. The submm-faint pBzK galaxies are tentatively detected at the $\sim
2\sigma$-level -- as also indicated by the slight increase in the azimuthally
averaged $S/N$ radial profile in Fig.\ \ref{figure:snr-laboca-stack}.  How do
these compare with previous submm stacking analyses of $K$-selected samples?

We find average 870-$\mu$m flux densities of $\rm 0.38\pm 0.04\,mJy$ and
$0.21\pm 0.10\,$mJy for our samples of submm-faint sBzK and pBzK galaxies.  In
comparison, Daddi et al.\ (2005) estimated an average 850-$\mu$m signal of
$\sim 0.82\,$mJy (corresponding to a 870-$\mu$m signal of $\sim
0.75\,$mJy\footnote{We scale 850-$\mu$m fluxes to 870-$\mu$m fluxes assuming an
optically thin, modified black body law with $\rm \beta = 1.5$, such that: $\rm
S_{870\mu m}/S_{850\mu m} = (850\mu m/870\mu m)^{2+\beta} = 0.92$.}) for a
sample of $\sim 100$ $K_{\mbox{\tiny{vega}}}\le 20$ submm-faint sBzK galaxies
within the SCUBA map of GOODS-N.  Takagi et al.\ (2007) reported an average
850-$\mu$m flux of $0.52\pm 0.19\,$mJy (corresponding to a 870-$\mu$m flux of
$0.48\pm 0.18\,$mJy) from a sample of 112 $K_{\mbox{\tiny{vega}}}\le 20$
sBzK galaxies selected within the part of the Subaru/XMM-Newton deep field
(SXDF) covered by the SCUBA HAlf Degree Extragalactic Survey (SHADES).
Finally, D08 stacked the 850-$\mu$m signal from 1421 $K_{\mbox{\tiny{vega}}}\ls
21.7$ sBzK galaxies and found $0.53\pm 0.06\,$mJy (corresponding to $0.49\pm
0.06\,$mJy at 870-$\mu$m). D08 also measured an average 850-$\mu$m flux of
$0.22\pm 0.18\,$mJy (or $0.20\pm 0.17\,$mJy at 870-$\mu$m) for a sample of 147
$K_{\mbox{\tiny{vega}}}\ls 21.7$ pBzK galaxies.

For the EROs and DRGs we find stacked 870-$\mu$m fluxes of $0.30\pm 0.03\,$mJy
and $0.32\pm 0.04\,$mJy, respectively.  Webb et al.\ (2004) used the Canada-UK
Deep Submillimeter Survey 03hr and 14hr fields (CUDSS03 and CUDSS14,
respectively) to perform a 850-$\mu$m stacking analysis of 164
$K_{\mbox{\tiny{vega}}}\ls 20.7$ EROs in the two fields, and found a stacked
signal for the entire ERO sample of $\langle S_{\mbox{\tiny{870$\mu$m}}}\rangle = 0.52\pm
0.09\,$mJy.  Similarly, Takagi et al.\ (2007) measured a stacked 870-$\mu$m
signal of $0.49\pm 0.16\,$mJy from a sample of 201 $K_{\mbox{\tiny{vega}}}\le
20$ EROs selected within SXDF/SHADES.  Turning to DRGs, Knudsen et al.\ (2005)
obtained a stacked 850-$\mu$m signal of $\rm 0.74\pm 0.24\,mJy$ from a sample
of 25 $K_{\mbox{\tiny{vega}}}\le 22.5$, submm-faint DRGs (uncorrected for
an average gravitational lens amplification of 20\,\%). Converting to a 870-$\mu$m
flux density and correcting for the lens amplification yields $0.54\pm 0.18\,$mJy.
Takagi et al.\ (2007) using significantly shallower SCUBA maps failed at detecting a
significant 850-$\mu$m signal from an average of 67 $K_{\mbox{\tiny{vega}}}\le
20$, submm-faint DRGs ($\rm \langle S_{870\mu m}\rangle = 0.39\pm 0.23\,mJy$).

We conclude that, within the errors, the previous stacking studies agree well
with our results. We also note that our study provide the first robust ($\ge
5$-$\sigma$) detection of submm-faint DRGs. 

From the stacked submm fluxes we are able to estimate average IR luminosities
and star formation rates (Table \ref{table:LIR-SFR}).  IR luminosities are
derived by adopting the IR-to-submm SED of Arp\,220 and scaling it to the
stacked submm fluxes (at the median redshifts derived from the redshift
distributions in \S~\ref{subsection:phot-z}) and integrating it from
$8-1000\,\mu$m. For comparison we also derive IR luminosities assuming that the
SEDs are described by modified black-body law with a dust temperature of
$T_{\mbox{\tiny{d}}}=35\,$K and $\beta = 1.5$, and integrating from $8-1000\,\mu$m. Star
formation rates ($SFR$) are derived following Kennicutt (1998): $SFR [\Msolar
\rm{yr}^{-1}] = 1.73\times 10^{-10}L_{\mbox{\tiny{IR}}}[\Lsolar]$.  This
conversion assumes a Salpeter initial mass function (Salpeter 1955). 
Of course, we stress that considerable uncertainty is associated with the derived
IR luminosities and $SFRs$ since they depend on the assumed SED and IMF.

The average IR luminosities and star formation rates estimated here for
$K_{\mbox{\tiny{vega}}}\le 20$ sBzK, ERO and DRG galaxies on the basis of their
stacked submm fluxes lie in the ranges $\sim 1-6\times 10^{11}\,\Lsolar$ and
$\sim 20-110\,\Msolar\,\rm{yr}^{-1}$, i.e.\ comparable to those of LIRGs and
ULIRGs in the local Universe. For the sBzK galaxies, the average IR luminosity
and star formation rate derived here are fully consistent with UV, 24-$\mu$m
and radio studies of these galaxies (Daddi et al.\ 2007).  We find that ERO and
DRG populations have significantly lower IR luminosities (by $\sim 40\%$) than
the sBzK galaxies. This is in part due to the fact that we have made no attempt
to weed out passive EROs/DRGs in the stacking analysis, and the stacked submm
flux from dusty, starforming EROs/DRGs is likely to be significantly higher.

\begin{deluxetable}{lccccccc}
\tabletypesize{\tiny}
\tablecaption{The average IR luminosities and star formation rates of BzK, ERO
and DRG galaxies, derived from their stacked 870-$\mu$m fluxes.}
\tablewidth{0pt}
\tablehead{
\colhead{Galaxy type}      & $\langle z\rangle$  & $L^{a}_{\mbox{\tiny{IR}}}$ (Arp\,220) & $SFR^b$ (Arp\,220)        & $L^{a}_{\mbox{\tiny{IR}}}$ ($T_{\mbox{\tiny{d}}}=35\,$K)  &  $SFR^b$ ($T_{\mbox{\tiny{d}}}=35\,$K) & $L^{a}_{\mbox{\tiny{IR}}}$ ($T_{\mbox{\tiny{d}}}=30\,$K)  &  $SFR^b$ ($T_{\mbox{\tiny{d}}}=30\,$K) \\ 
                           &                     & [$\times 10^{11}\Lsolar$]             & [$\Msolar\,\rm{yr}^{-1}$  & [$\times 10^{11}\Lsolar$]                                 &  [$\Msolar\rm{yr}^{-1}$]               & [$\times 10^{11}\Lsolar$]                                 &  [$\Msolar/\rm{yr}^{-1}$]} 
\startdata
sBzK                       & 1.8                 & $6.3\pm 0.4$                          & $109\pm 7$                & $3.7\pm 0.3$                                              &  $65\pm 5$                             & $2.1\pm 0.2$                                              &  $37\pm 3$\\
pBzK                       & 1.6                 & $2.9\pm 1.0$                          & $50\pm 17$                & $1.7\pm 0.7$                                              &  $30\pm 11$                            & $1.0\pm 0.4$                                              &  $17\pm 6$\\
ERO                        & 1.3                 & $3.6\pm 0.3$                          & $62\pm 5$                 & $2.3\pm 0.2$                                              &  $39\pm 3$                             & $1.3\pm 0.1$                                              &  $22\pm 2$\\
DRG                        & 1.4                 & $3.6\pm 0.4$                          & $62\pm 7$                 & $2.3\pm 0.3$                                              &  $40\pm 4$                             & $1.3\pm 0.1$                                              &  $22\pm 3$\\
\hline
sBzK~($z < 1.4$)           & 1.1                 & $3.9\pm 0.8$                          & $67\pm 14$                & $2.4\pm 0.5$                                              &  $42\pm 9$                             & $1.3\pm 0.3$                                              &  $10\pm 2$\\
sBzK~($z > 1.4$)           & 2.0                 & $5.8\pm 0.5$                          & $101\pm 9$                & $3.9\pm 0.3$                                              &  $67\pm 6$                             & $2.3\pm 0.2$                                              &  $39\pm 4$\\
pBzK~($z < 1.4$)           & 1.2                 & $. . .$                               & $. . .$                   & $. . .$                                                   &  $. . .$                               & $. . .$                                                   &  $. . .$\\
pBzK~($z > 1.4$)           & 1.9                 & $4.7\pm 1.2$                          & $82\pm 21$                & $3.0\pm 0.8$                                              &  $52\pm 14$                            & $1.7\pm 0.5$                                              &  $30\pm 8$\\
ERO~($z < 1.4$)            & 1.1                 & $2.5\pm 0.4$                          & $43\pm 7$                 & $1.5\pm 0.2$                                              &  $26\pm 4$                             & $0.8\pm 0.1$                                              &  $14\pm 2$\\
ERO~($z > 1.4$)            & 2.1                 & $5.8\pm 0.5$                          & $101\pm 9$                & $3.7\pm 0.3$                                              &  $64\pm 6$                             & $2.2\pm 0.2$                                              &  $37\pm 3$\\
DRG~($z < 1.4$)            & 1.1                 & $2.8\pm 0.6$                          & $48\pm 10$                & $1.7\pm 0.3$                                              &  $29\pm 6$                             & $0.9\pm 0.2$                                              &  $16\pm 3$\\
DRG~($z > 1.4$)            & 2.4                 & $5.6\pm 0.6$                          & $96\pm 10$                & $3.8\pm 0.4$                                              &  $67\pm 7$                             & $2.3\pm 0.2$                                              &  $40\pm 4$\\
\enddata
\tablenotetext{~}{$^{a}$IR luminosities are obtained by integrating the SED over the wavelength range $8-1000\,\mu$m.}
\tablenotetext{~}{$^{b}$Star formation rates a derived using  $SFR [\Msolar \rm{yr}^{-1}] = 1.73\times 10^{-10}L_{\mbox{\tiny{IR}}}[\Lsolar]$ (Kennicutt 1998).}
\label{table:LIR-SFR}
\end{deluxetable}

\subsection{Stacking in redshift bins}\label{subsection:stacking-z}
\begin{figure*}[t] 
\includegraphics[angle=0,width=0.5\hsize]{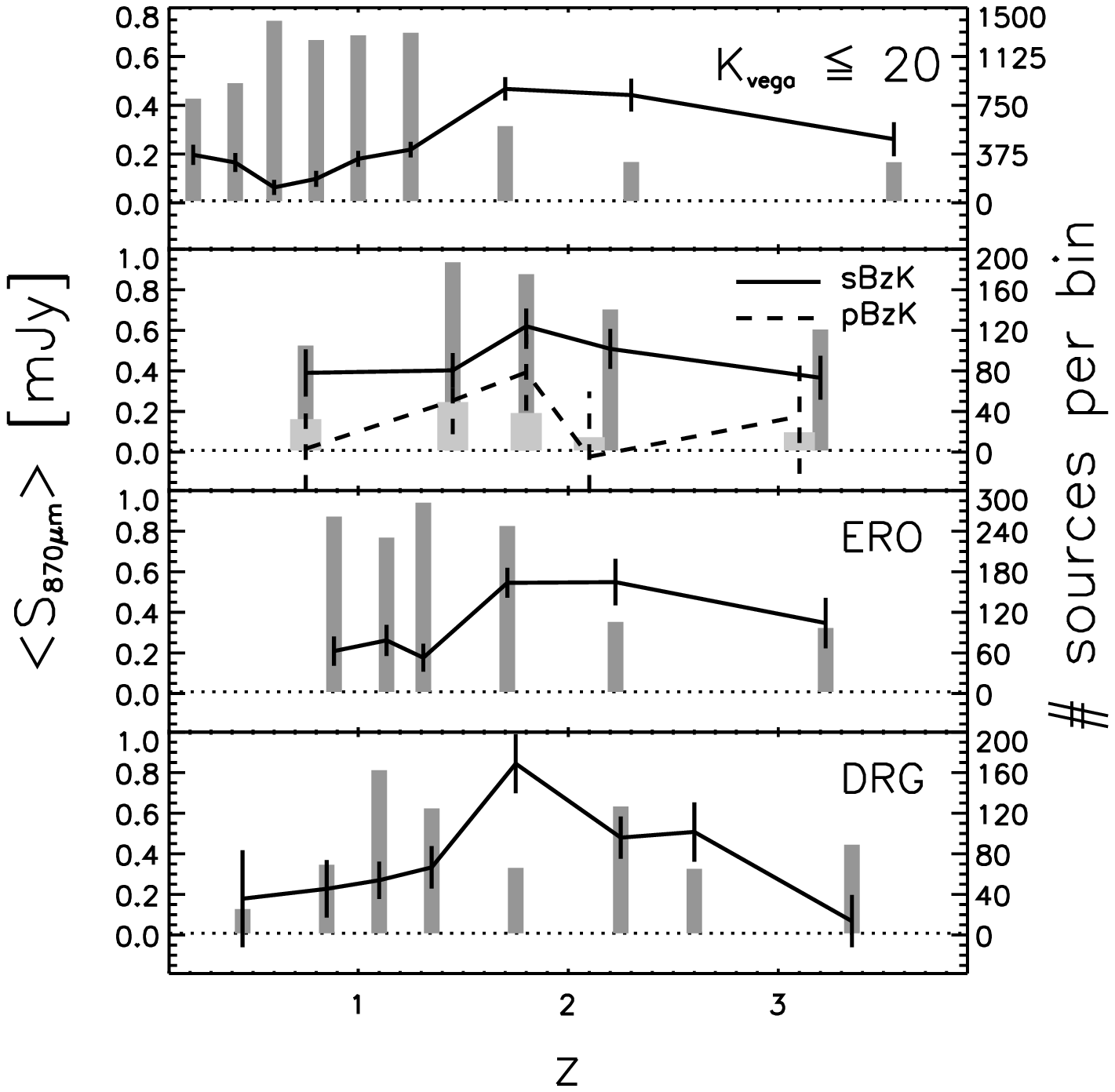}\includegraphics[angle=0,width=0.5\hsize]{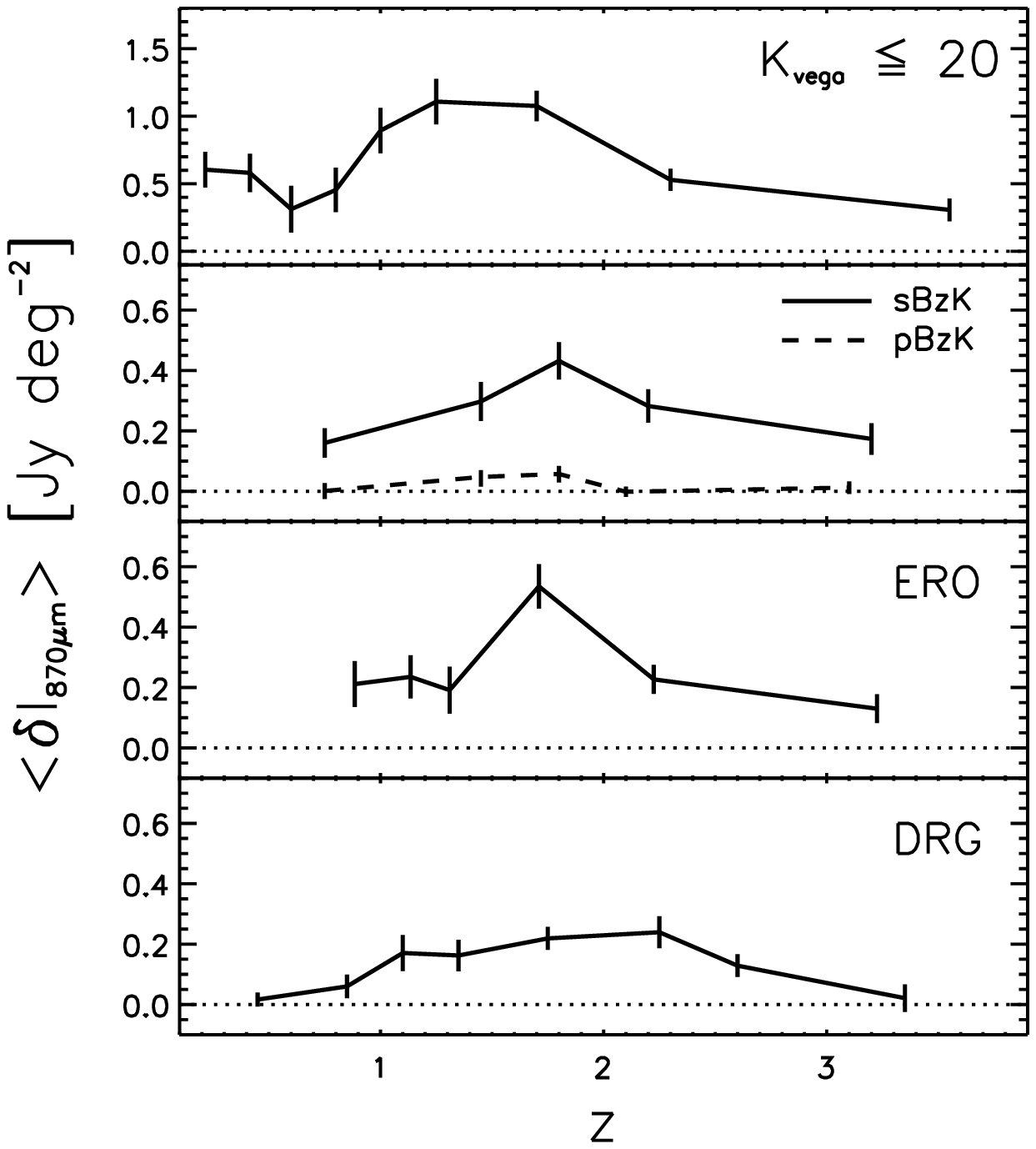} 
\caption{Stacked 870-$\mu$m flux densities (left panels) and the corresponding
contribution to the 870-$\mu$m extragalactic background light (right panels) as
a function of redshift for the $K_{\mbox{\tiny{vega}}}\le 20$, BzK, ERO and DRG samples. The gray histograms in
the left hand panels indicate the number of sources in each bin (corresponding y-axis is
on the right hand side). Broader, more light-grey histograms have been used for the pBzK galaxies.
}
\label{figure:stack870_z} 
\end{figure*}

Using the photometric redshifts obtained in \S~\ref{subsection:phot-z} we can
stack our samples into separate redshift bins, thereby allowing us to determine
which redshifts are contributing the most to the stacked submm signals. Redshift bins were
chosen so that they were larger than the typical redshift uncertainty, and
provided roughly the same number of sources in each bin. The latter ensured
that the same sensitivity was reached in each bin, thus allowing for a direct
comparison.  Fig.\ \ref{figure:stack870_z} shows the average flux densities of
the different samples as a function of redshift.  We stress that due to the
essentially flat selection function of submm surveys over the redshift range
$1\ls z\ls 8$ (Blain \& Longair 1993), the comparison of stacked submm fluxes
at different redshifts directly translates into a comparison between far-IR
luminosities (and thus star formation rates -- Kennicutt 1998) between these
redshift bins (to the extent that submm flux is a measure of far-IR
luminosity).

The average submm signal from $K_{\mbox{\tiny{vega}}}\le 20$ galaxies is found
to be roughly constant ($\sim 0.1-0.2\,$mJy) out to $z\sim 1.4$, and consistent
with the average flux density of the full sample (Table \ref{table:stacking}).
At $z\sim 1.7$, however, the submm signal has increased to $\sim 0.4\,$mJy. The
stacked submm fluxes of $z < 1.4$ and $z > 1.4$ $K_{\mbox{\tiny{vega}}}\le 20$
sources are $0.15\pm 0.01\,$mJy and $0.47\pm 0.03\,$mJy, respectively (Table
\ref{table:stacking}).  A similar increase in the average submm flux at $z >
1.4$ is seen for the ERO and DRG populations. The average IR luminosities and
star formation rate for $z
> 1.4$ EROs ($L_{\mbox{\tiny{IR}}}\simeq 4.9\times 10^{11}\,\Lsolar$ and $SFR
\simeq 85\,\Msolar\,\rm{yr}^{-1}$) are about $3\times$ higher than for $z <
1.4$ EROs ($L_{\mbox{\tiny{IR}}}\simeq 1.8\times 10^{11}\,\Lsolar$ and $SFR
\simeq 31\,\Msolar\,\rm{yr}^{-1}$).  For DRGs the difference is about a factor
of two: $z > 1.4$ DRGs have on average $L_{\mbox{\tiny{IR}}}\simeq 4.9\times
10^{11}\,\Lsolar$ and $SFR \simeq 85\,\Msolar\,\rm{yr}^{-1}$ while $z < 1.4$
DRGs have $L_{\mbox{\tiny{IR}}}\simeq 2.4\times 10^{11}\,\Lsolar$ and $SFR
\simeq 42\,\Msolar\,\rm{yr}^{-1}$).  These findings also fit with the ERO and
DRG redshift distributions (Fig.\ \ref{figure:Nz}), where we found evidence for
two sub-populations separated at $z\sim 1.6$.  Due to our magnitude cut-off at
$K_{\mbox{\tiny{vega}}}\le 20$, our samples are biased towards increasingly
more massive galaxies at higher redshifts. The above results, therefore,
suggest that the star formation activity in massive galaxies increase with
redshift, and that a dominant fraction of massive EROs/DRGs at $z>1.4$ are
actively starforming galaxies, and not old, red galaxies. The less massive EROs
and DRGs, which aren't picked up at high-$z$ given the
$K_{\mbox{\tiny{vega}}}\le 20$ limit, have lower star formation rates, and
possibly a higher fraction of evolved, passive galaxies, than the more massive
EROs/DRGs.  The gradual drop in the average submm signals at $z>2.5$ is likely
a reflection of the incompleteness in our $K$-band selection at these
redshifts.

Turning to the BzK galaxies, we find that sBzKs exhibit a positive and, within
the error bars, constant 870-$\mu$m signal ($\sim 0.5\,$mJy) over the redshift
range $1\ls z \ls 3$.  This supports the notion that the sBzK-criterion selects
star forming galaxies across this redshift range, and that for our $K$-band
magnitude limit of $K_{\mbox{\tiny{vega}}}\le 20$, the distribution of star
formation rates of sBzK galaxies is roughly constant within this redshift
range. In contrast, the pBzK galaxies show no evidence of significant submm
signal in any redshift bin, which is consistent with these galaxies being
devoid of significant star formation.  D08 reported a 850-$\mu$m signal of
$0.89\pm 0.34$\,mJy ($\sim 2.6$-$\sigma$) for pBzKs at $z<1.4$ but no
significant signal for pBzKs at $z>1.4$. They argued that the submm signal for
$z<1.4$ pBzK galaxies was due to contamination by star forming galaxies at
$z<1.4$.  In comparison, we find no significant stacked 870-$\mu$m signal
($-0.14\pm 0.17\,$mJy) for $z<1.4$ pBzKs and only a marginal signal for $z>1.4$
pBzKs ($0.36\pm 0.12\,$mJy).

\subsection{Stacking in 24-$\mu$m bins}
Using the 24-$\mu$m source catalogue from the FIDEL survey (which includes all
sources $> 27\,\mu$Jy, 5-$\sigma$ point-source sensitivity -- Dickinson et al.\
in prep), a total of 466/53, 511 and 400 sources in the sBzK/pBzK, ERO and DRG
samples, respectively, were identified at 24-$\mu$m. Their average submm fluxes
are given in Table \ref{table:stacking} along with those of the 24-$\mu$m faint
($ < 27\,\mu$Jy) subsets. The 24-$\mu$m detected subsets have $\gs 5\times$
higher average submm flux densities than the 24-$\mu$m faint sources.  This is
not surprising since the mid-IR is known to trace the thermal dust emission, and
one might even expect a correlation between the submm and mid-IR emission.

\begin{figure*}[t] 
\includegraphics[angle=0,width=0.5\hsize]{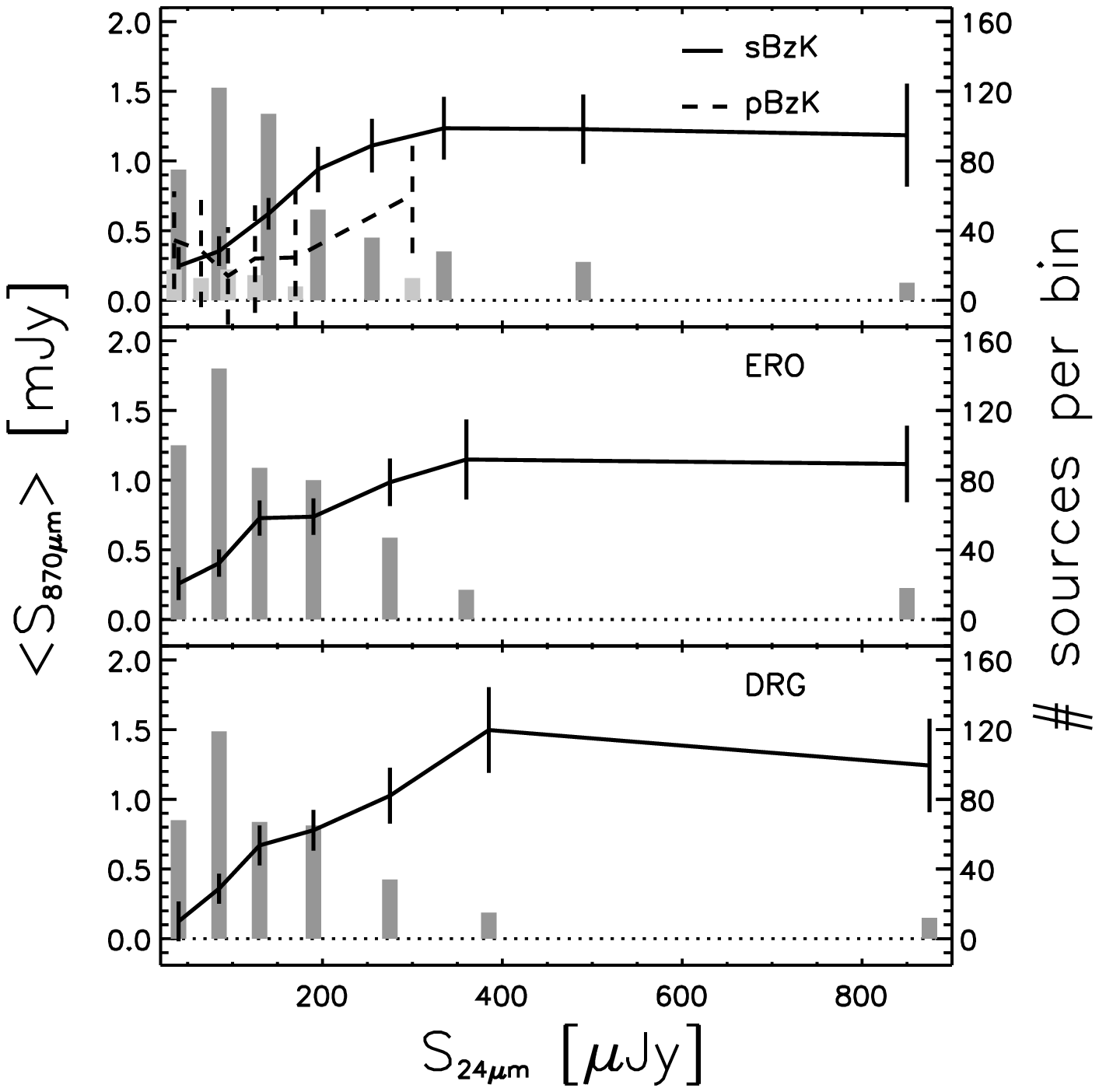} 
\includegraphics[angle=0,width=0.5\hsize]{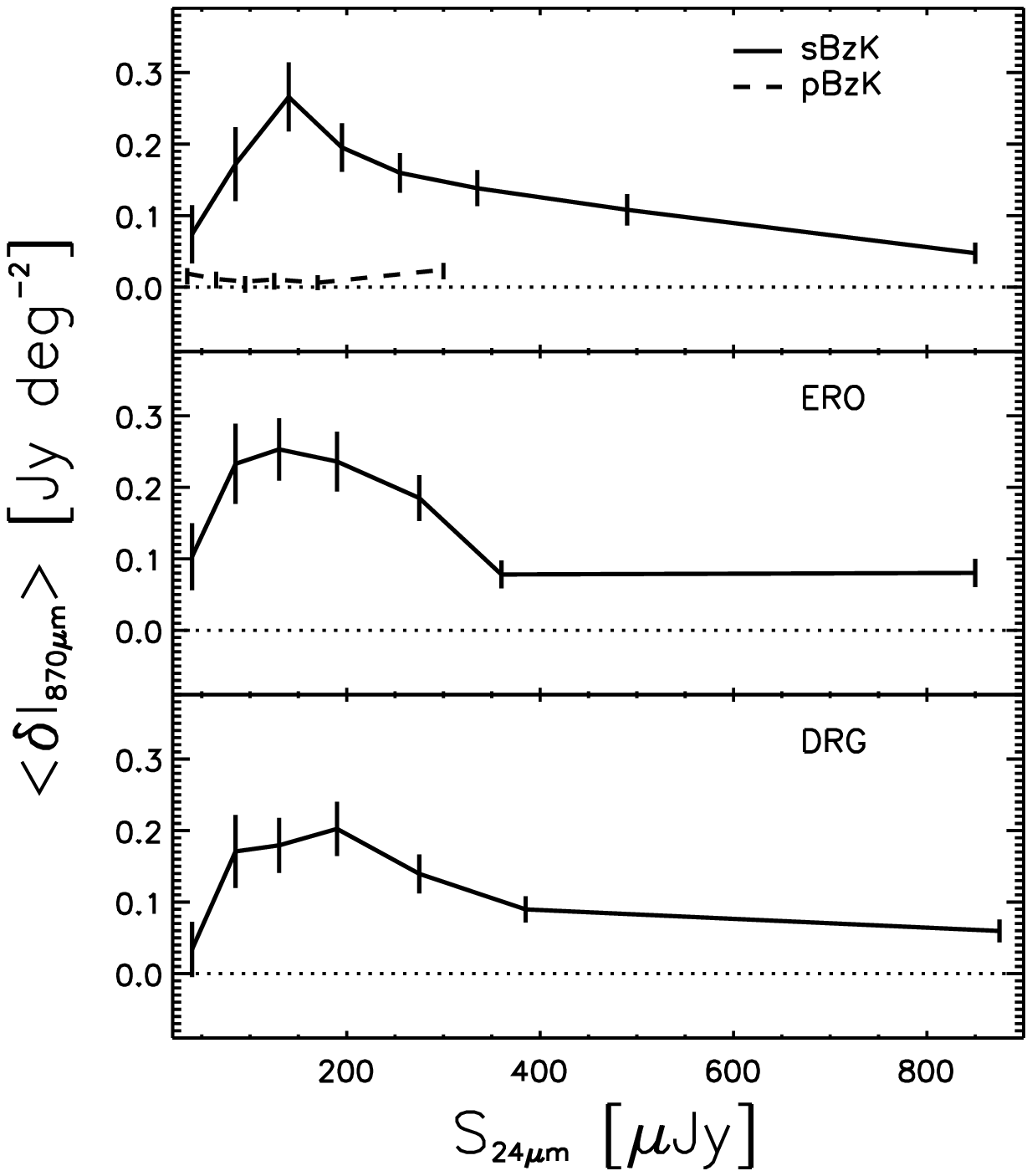} 
\caption{Stacked 870-$\mu$m flux densities (left panels) and the corresponding
contribution to the 870-$\mu$m extragalactic background light (right panels) as
a function of 24-$\mu$m flux density for the BzK, ERO and DRG samples. The gray
histograms in the left hand panels indicate the number of sources in each bin
(corresponding y-axis is on the right hand side). Broader, more light-grey
histograms have been used for the pBzK galaxies.  An all cases, a significant
$S_{\mbox{\tiny{870$\mu$m}}}-S_{\mbox{\tiny{24$\mu$m}}}$ correlation is found for
$S_{\mbox{\tiny{24$\mu$m}}}\ls 350\,\mu$Jy, indicating that across this flux
range, 24-$\mu$m observations trace star formation.  For
$S_{\mbox{\tiny{24$\mu$m}}}\gs 350\,\mu$Jy, however, the relation flattens,
suggestive of an increase in the fraction of AGN-dominated systems.
}
\label{figure:stack870_24um} 
\end{figure*}

In order to investigate the latter, we measured the stacked 870-$\mu$m signal
as a function of 24-$\mu$m flux density bins (Fig.\
\ref{figure:stack870_24um}).  The sBzK, ERO and DRG galaxies all exhibit a
similar behaviour, namely a significant linear correlation between the stacked
870-$\mu$m and 24-$\mu$m flux density up to $S_{\mbox{\tiny{24$\mu$m}}} \simeq
350\,\mu$Jy, given by $\langle S_{\mbox{\tiny{870$\mu$m}}}\rangle = 4.5\times
10^{-3}\langle S_{\mbox{\tiny{24$\mu$m}}}\rangle$, followed by a flattening of
the relation for $S_{\mbox{\tiny{24$\mu$m}}} > 350\,\mu$Jy.  Not surprisingly,
the pBzKs are not significantly detected at 870-$\mu$m in any of the 24-$\mu$m
bins, and no submm-mid-IR correlation is seen.

Although the 24-$\mu$m selection function (Daddi et al.\ 2007) depends strongly
on redshift (unlike the selection function at submm wavelengths), the observed
$S_{\mbox{\tiny{24$\mu$m}}}-S_{\mbox{\tiny{870$\mu$m}}}$ correlation strongly
suggests that 24-$\mu$m measurements (with $S_{\mbox{\tiny{24$\mu$mJy}}}\gs
350\,\mu$Jy) trace systems dominated by star formation, and can be used to
derive IR luminosities and star formation rates.
The constant $S_{\mbox{\tiny{870$\mu$m}}}\simeq 1.3\,$mJy ratio
for $S_{\mbox{\tiny{24$\mu$m}}}\simeq 350-1000\,\mu$Jy is
likely to reflect significant contamination by an AGN at these high 24-$\mu$m
flux densities.  Since the mid-IR will be more sensitive to warm dust
($T_{\mbox{\tiny{d}}}\sim 80-200\,$K) than the submm, which largely traces cold
dust ($T_{\mbox{\tiny{d}}}\sim 20-60\,$K), the ability of the observed
24-$\mu$m emission to reliably trace IR luminosity and star formation may be
compromised by AGN-heated warm dust.  The turn-over from a linear to a flat
$S_{\mbox{\tiny{870$\mu$m}}} - S_{\mbox{\tiny{24$\mu$m}}}$ relation occurs at
$S_{\mbox{\tiny{24$\mu$m}}}\simeq 350\,\mu$Jy, which corresponds to
$L_{\mbox{\tiny{IR}}}< 1.5\times 10^{12}\,\Lsolar$ at $z\sim 2$.

Our findings are in line with those of Papovich et al.\ (2007), who found that
IR luminosities derived from 24-, 70-, and 160-$\mu$m {\it Spitzer} data (of a
sample of $K$-selected galaxies at $1.5\ls z\ls 2.5$ with
$S_{\mbox{\tiny{24$\mu$mJy}}}\simeq 50-250\,\mu$Jy) were in good agreement with
those derived from 24-$\mu$m data alone.  For sources with
$S_{\mbox{\tiny{24$\mu$mJy}}} > 250\,\mu$Jy, however, the latter would be $\sim
2-10\times$ higher, suggesting that the AGN may contribute significantly to the
high 24-$\mu$m emission.  Similarly, Daddi et al.\ (2004, 2005) found that a
large fraction ($> 30\%$) of sBzK galaxies with $L_{\mbox{\tiny{IR}}}\gs 1.5\times
10^{12}\,\Lsolar$ (which corresponds to the IR luminosity where we find a
turn-over in the $S_{\mbox{\tiny{870$\mu$m}}} - S_{\mbox{\tiny{24$\mu$m}}}$
relation) show an excess of emission in the near-IR (rest-frame) and are
statistically detected in hard X-rays -- evidence of powerful AGN in these
very IR-luminous systems.

\subsection{Contributions to the extragalactic background light}\label{subsection:analysis-EBL}
Turning to the contribution to the submm extragalactic background light (EBL)
by the different populations, we adopt the spectral approximations to the EBL
at submm wavelengths from COBE/FIRAS (Puget et al.\ 1996; Fixsen et al.\ 1998),
including their uncertainties. In doing so we adopt a value of the EBL at
870-$\mu$m of $44\pm 15\,\rm{Jy}\,\rm{deg}^{-2}$.  From the surface densities
of the samples we derive their contributions to the 870-$\mu$m extragalactic
background light (Table \ref{table:stacking}).  We find that the total
contribution from all $K_{\mbox{\tiny{vega}}}\le 20$ sources to the 870-$\mu$m
EBL is $6.61\pm 0.34\,$Jy$\,$deg$^{-2}$, or $15.0\pm 5.2\,\%$.  For the BzK
galaxies we find that the passive ones contribute $\ls  1\,\%$ to the
870-$\mu$m EBL, while the star forming population contribute $1.70\pm
0.24\,$Jy$\,$deg$^{-2}$, corresponding to $3.9\pm 1.4\,\%$.  The EROs
contribute a similar amount to the EBL at 870-$\mu$m as the sBzK galaxies
($1.85\pm 0.20\,$Jy$\,$deg$^{-2}$ or $4.2\pm 1.5\,\%$), while the contribution
from DRGs, owing to their lower surface density, is about 80\% smaller
($1.09\pm 0.18\,$Jy$\,$deg$^{-2}$ or $2.5\pm 0.9\,\%$). The combined
BzK/ERO/DRG sample, which includes the bulk of massive starforming galaxies at
$z\gs 1$, contribute with $8.4\pm 2.9\,\%$ to the EBL at 870-$\mu$m after
accounting for overlap between the populations.

According to Takagi et al.\ (2007) sBzK galaxies down to
$K_{\mbox{\tiny{vega}}}\ls 20$ contribute with $3.8\pm 1.2\,$Jy$\,$deg$^{-2}$
(or $8.3\pm 3.9\,\%$) to the background light at 850-$\mu$m ($46\pm
16\,$Jy$\,$deg$^{-2}$  -- Puget et al.\ 1996; Fixsen et al.\ 1998).  Webb et
al.\ (2004) found that the ERO population down to $K_{\mbox{\tiny{vega}}}<20.7$
constitutes 7-11\% of the EBL at 850-$\mu$m, while Takagi et al.\ (2007)
reported a 850-$\mu$m EBL contribution from $K_{\mbox{\tiny{vega}}}\ls 20$ EROs
of $5.1\pm 1.5\,$Jy$\,$deg$^{-2}$ ($11\pm 5\,\%$) to the total background at
850-$\mu$m.  Knudsen et al.\ (2005) found that DRG galaxies down to
$K_{\mbox{\tiny{vega}}}<22.5$ contribute $\sim 7.7\,$Jy$\,$deg$^{-2}$ ($\sim
17\,\%$) of the EBL at 850-$\mu$m.

\bigskip

Next, let us look at the contribution to the EBL at 870-$\mu$m from the
different samples as a function of redshift and 24-$\mu$m flux bins (right
panels in Fig.\ \ref{figure:stack870_z} \& \ref{figure:stack870_24um}).  We
calculate the contributions using the stacked submm flux and the surface
density of sources in each redshift and 24-$\mu$m bin.  

Looking at the redshift dependence first, we see that the strongest
contribution to the 870-$\mu$m EBL by $K_{\mbox{\tiny{vega}}}\le 20$ galaxies
in a given redshift bin, is coming from sources in the redshift range $1 < z <
2$ -- they contribute with $7.3\pm 0.2\,\rm{Jy}\,\rm{deg}^{-2}$ to the submm
EBL, which corresponds to $\sim 50\,\%$ of the total contribution from
$K_{\mbox{\tiny{vega}}}\le 20$ galaxies (Table \ref{table:stacking}).  Although
$K_{\mbox{\tiny{vega}}}\le 20$ galaxies at $z > 2$ have significant submm
emission (\S~\ref{subsection:stacking-z}), their low abundance means that they
contribute $\ls 1\%$ to the observed submm EBL.  In contrast, $z < 1$
$K_{\mbox{\tiny{vega}}}\le 20$ galaxies are abundant (Fig.\ \ref{figure:Nz}),
but their low average submm fluxes implies a $\ls 1\,\%$ contribution to the
submm EBL.

The bulk ($\sim 80\,\%$) of the contribution to the submm EBL from 24-$\mu$m detected
(i.e.\ $S_{\mbox{\tiny{24$\mu$Jy}}} > 27\,\mu$Jy) sBzK, ERO and DRG galaxies comes from sources with
$S_{\mbox{\tiny{24$\mu$Jy}}} \simeq 50-350\,\mu$Jy, with only a minor fraction
coming from brighter, presumably, AGN-dominated sources.

\subsection{Stacking across the BzK and RJK diagrams}
The sBzK and pBzK selections are designed to locate star forming and passive
galaxies in the redshift range $1.4 \le z \le 2.5$, while the ERO and DRG
colour criteria are designed to select extremely red (either due to dust
extinction or old age) galaxies at $z > 1$. Still, the exact definitions of
these colour criteria are somewhat arbitrary. Given the large number of sources
available to us, we are in a position to construct stacks of the 870-$\mu$m
signal from statistically significant subsets of galaxies across the BzK and
RJK diagrams, thus allowing us to see where in these colour-colour diagrams the
submm signal is coming from.  Ultimately, this may allow us to fine-tune the
sBzK/pBzK criteria, as well as potentially identify regions of the RJK-diagram
containing dusty/starforming vs.\ old/passive EROs and DRGs. 
\begin{figure*}[t] 
\includegraphics[angle=0,width=1.0\hsize]{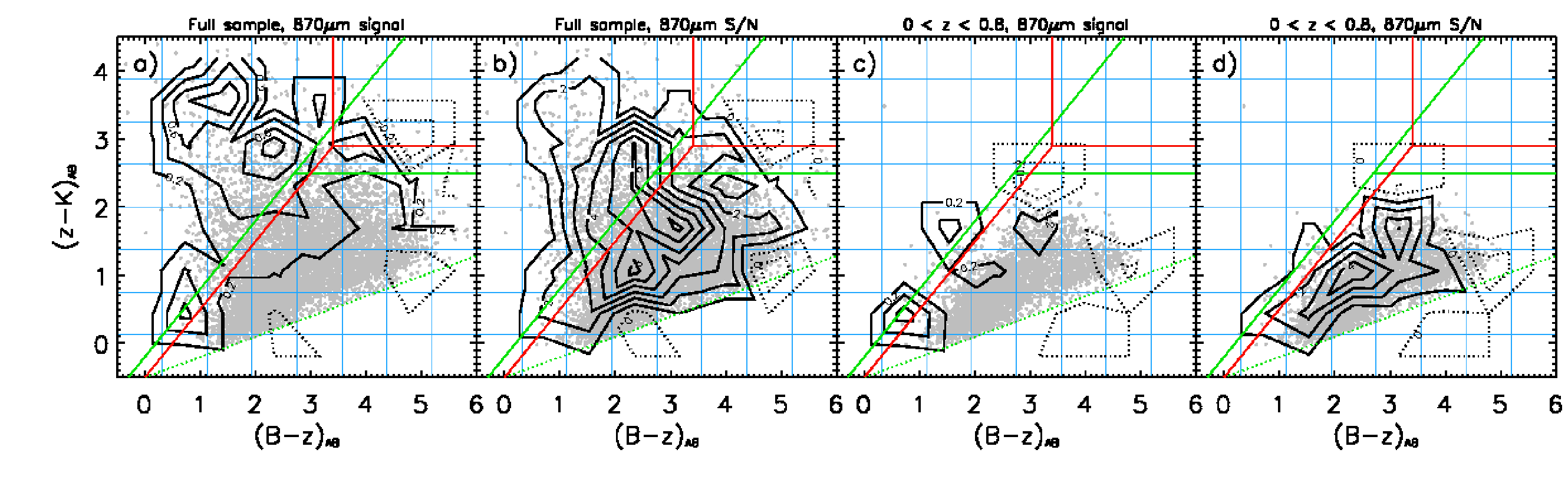}
\includegraphics[angle=0,width=1.0\hsize]{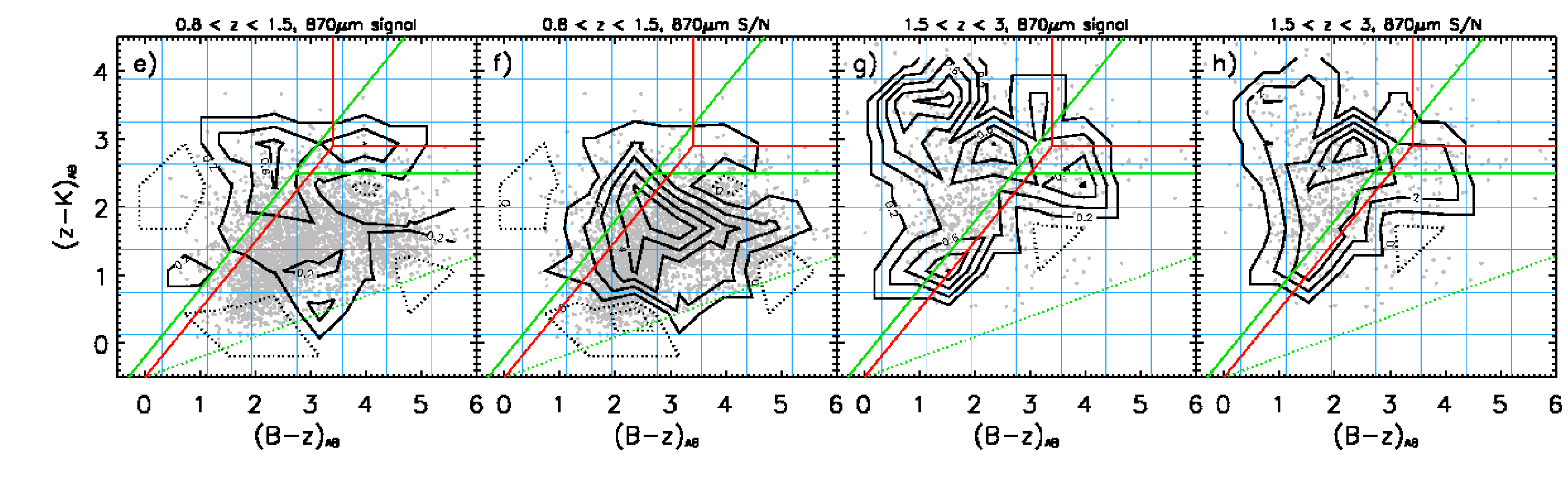} 
\caption{Contours of the stacked 870-$\mu$m signal (in units of mJy) and $S/N$ across the BzK
diagram, obtained by stacking the submm flux in a
regular grid (shown in blue). We have stacked the entire sample within the full redshift range,
as well as subsets within the redshift ranges $z=0-0.8$, $0.8-1.5$, and $1.5-3.0$.
The sBzK- and pBzK-criteria by Daddi et al.\ (2004) are shown as solid green lines, while
the refined selection criteria proposed in this paper are shown in red.
The contours clearly supports the BzK-selection technique as the submm signal is seen to
be almost entirely dominated by sBzK galaxies in the redshift range $z=1.5-3$.
}
\label{figure:BzK_stack} 
\end{figure*}
\begin{figure*}[t] 
\includegraphics[angle=0,width=1.0\hsize]{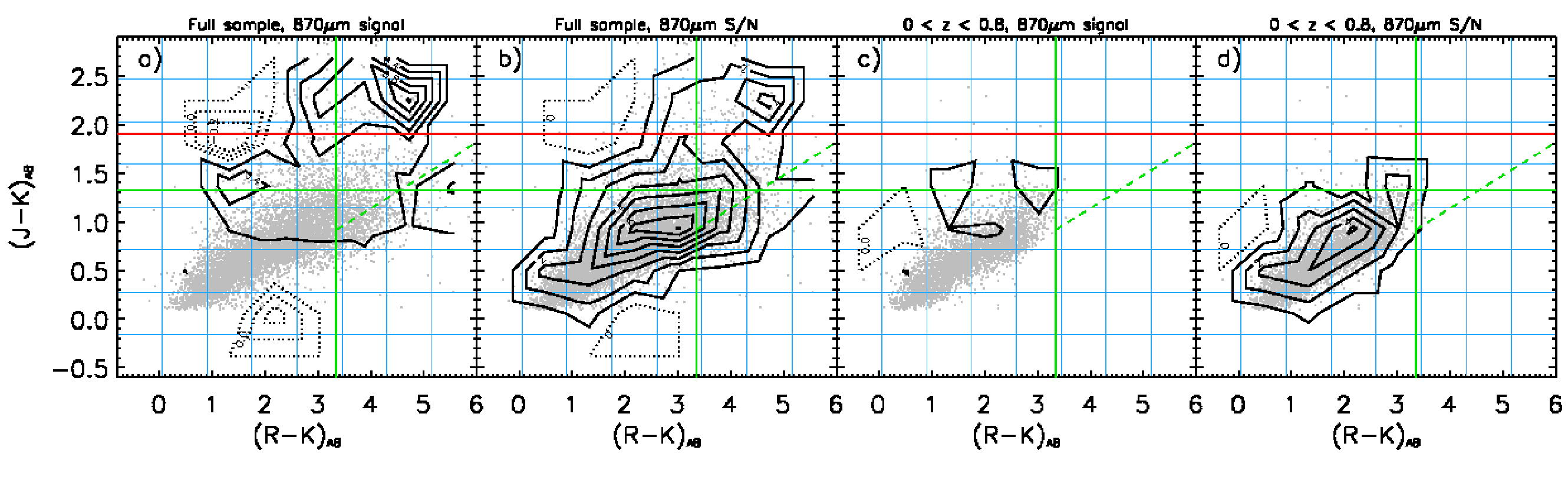}
\includegraphics[angle=0,width=1.0\hsize]{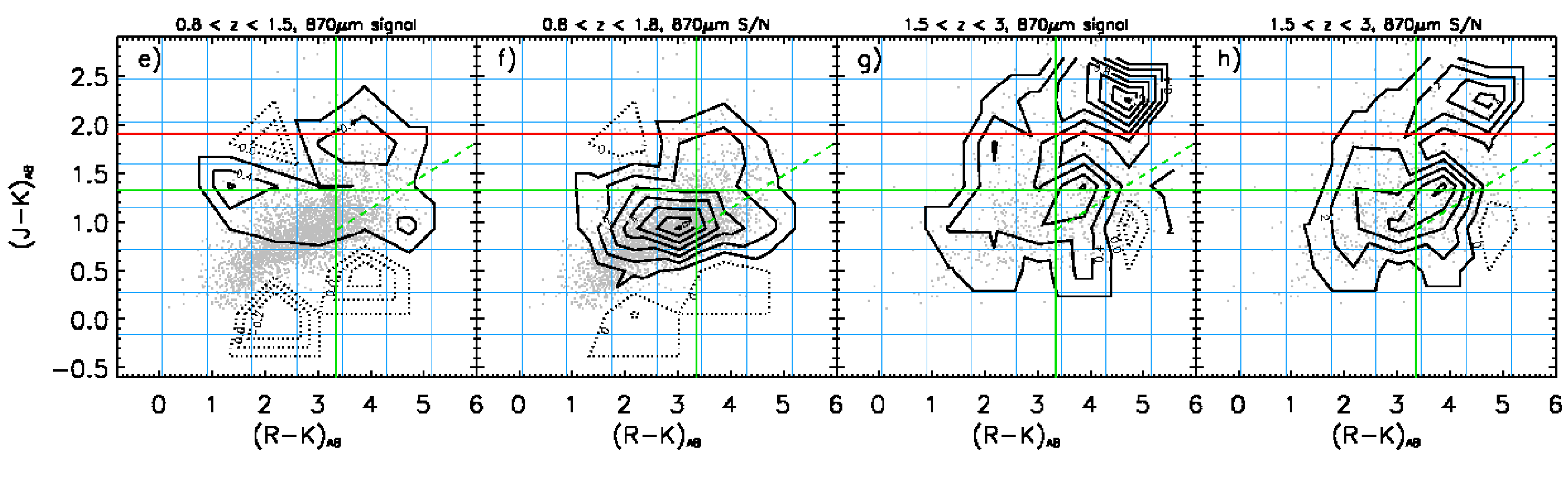} 
\caption{Contours of the stacked 870-$\mu$m signal (in units of mJy) and $S/N$ across the RJK
diagram, obtained by stacking the submm flux in a
regular grid (shown in blue). We have stacked the entire sample within the full redshift range,
as well as subsets within the redshift ranges $z=0-0.8$, $0.8-1.5$, and $1.5-3.0$.
The ERO- and DRG-criteria are shown as vertical and horizontal solid green lines, while
the starforming vs.\ passive ERO criterion proposed by Pozzetti \& Mannucci (2000) 
is illustrated by the green dashed line (with starforming and passive EROs lying above
and below the line, respectively). The strongest submm signal is coming from galaxies in the
redshift range $z=1.5-3$ lying in the DRG-ERO overlap region.}
\label{figure:RJK_stack} 
\end{figure*}

In Fig.\ \ref{figure:BzK_stack} and \ref{figure:RJK_stack} we show the signal- and $S/N$-contours of the stacked
870-$\mu$m signal obtained across the BzK- and RJK-diagrams, where the 
stacking has been carried out within equally sized grid-cells. We have done this
for the full sample, as well as for subsets of the sample within the redshift intervals
$0 < z < 0.8$, $0.8 < z < 1.5$, and $1.5 < z < 3.0$. These intervals were motivated by the
redshift distributions in Fig.\ \ref{figure:Nz}, which showed evidence of distinct populations
between $0.8 < z < 1.5$ and $1.5 < z < 3.0$.

Considering Fig.\ \ref{figure:BzK_stack}a first, we see that the sBzK criterion by Daddi et
al.\ (2004) seems to be robust in the sense that the bulk of the 870-$\mu$m
signal is emerging from the sBzK region.  Comparing with Fig.\
\ref{figure:cc-diagram}a indicates that virtually the entire submm signal is
coming from the subset of sBzK galaxies (with a slight overlap into the pBzK
and non-BzK regions), which have also been classified as EROs.
Due to their extremely red colours and relatively strong submm signal, these galaxies
are very likely to be amongst the most dusty, star forming sources of the 
submm-faint $K_{\mbox{\tiny{vega}}}\le 20$ selected galaxies, yet their brightness ensures
that they are detected in the blue (thus qualifying as sBzK galaxies).

The submm signal clearly extends into the pBzK region suggesting
that some contamination by star forming galaxies occurs. This is partly due the
low density of sources in the pBzK-region which implies that more sources will
scatter into the region (due to photometric errors) than out of it. 
Based on the 870-$\mu$m signal and $S/N$ contours in Fig.\ \ref{figure:BzK_stack},
we propose a refinement of the sBzK and pBzK criteria, namely
\begin{eqnarray}
(z-K)_{\mbox{\tiny{AB}}} &\ge& (B-z)_{\mbox{\tiny{AB}}} -0.5\\ 
(B-z)_{\mbox{\tiny{AB}}} &\le& 3.4
\end{eqnarray}
for sBzKs, and
\begin{eqnarray}
(z-K)_{\mbox{\tiny{AB}}} &\ge& 2.9\\
(B-z)_{\mbox{\tiny{AB}}} &>& 3.4
\end{eqnarray}
for pBzKs. 

That these refined selection criteria might do a better job at selecting starforming and passive 
BzK galaxies, is confirmed if we consider how the submm signal across the BzK-diagram changes with
redshift (Fig.\ \ref{figure:BzK_stack}c-h). We find that only marginal 870-$\mu$m emission
emerges from galaxies in the redshift interval $0 < z < 0.8$, while in the $0.8 < z
< 1.5$ interval, significant submm emission ($\sim 0.6\,$mJy) starts to appear from galaxies lying
within the sBzK region. At $1.5 < z < 3$ a strong submm signal ($\sim 1\,$mJy) and nearly all
of the submm emission is coming from the sBzK region. This observed migration of the stacked submm signal
towards the sBzK region as a function of redshift, illustrates the ability of the
sBzK-criterion to select starforming galaxies at $z\ge 1.5$.

Turning to the RJK-diagram, the bulk of the submm-signal is found to come from
the overlap region between DRGs and EROs, in particular from ERO/DRG galaxies
with $(J-K)_{\mbox{\tiny{AB}}} > 1.9$. The latter, together with the tentative
evidence of an underdensity of sources at $(J-K)_{\mbox{\tiny{AB}}} = 1.9$ (see
Fig.\ \ref{figure:cc-diagram}), suggests that $(J-K)_{\mbox{\tiny{AB}}} >
1.9$ may be a useful criterion for selecting a dusty, starforming galaxies.
This is strengthened by the fact that the submm signal becomes even stronger
and further concentrated towards the ERO/DRG overlap region if we consider only
the subset of sources at $z=1.5-3$ (Fig.\ \ref{figure:RJK_stack}g-h).
In the redshift range $0.8 < z < 1.5$, the submm signal is much weaker ($\sim 0.6\,$mJy) 
and comes from ERO/DRG galaxies with $(J-K)_{\mbox{\tiny{AB}}} < 1.9$. 

It is well-known that the $K_{\mbox{\tiny{vega}}} \le
20$ ERO population is a heterogeneous population, consisting of roughly a
fifty-fifty mix of dusty, star forming EROs and evolved, passive EROs (Spinrad
et al.\ 1999; Dey et al.\ 1999; Cimatti et al.\ 1999; Mannucci et al.\ 2002).
Pozzetti \& Mannucci (2000) argued that a crude separation between the two
types of EROs could be made based on the criterion $(J-K)_{\mbox{\tiny{AB}}} =
0.34 (R-K)_{\mbox{\tiny{AB}}} - 0.22$ (and $(R-K_S)_{\mbox{\tiny{AB}}} \ge
3.35$)\footnote{We have converted the Pozzetti \& Mannucci criterion, which was
in the vega system, into AB magnitudes.}, with dusty, star forming EROs lying
above the relation and old, passive EROs below it. While our stacking analysis
lends some merit to the Pozzetti \& Mannucci criterion, as the bulk of the
submm signal is clearly found above it, significant submm emission is also
detected in the passive ERO region (in particular for sources at $z > 1.5$),
suggesting that blindly applying the criterion does not produce clean samples of starforming
and passive EROs (see also Smail et al.\ 2002b).

Finally, we caution that by adopting equally sized grid cells, some cells will
contain a significantly larger number of sources than others, thereby
introducing a potential skewing of the measured $S/N$ across the diagrams (which
is why we also show the variation of the average submm flux density across the diagrams).  As
a check on our results, we therefore adopted two alternative methods for binning
the sources. First, an adaptive mesh was applied by
requiring that no more than 200 sources were allowed within a given cell, and, secondly,
the 50 nearest neighbours of each source were identified and stacked.  
Reassuringly, the two additional binning methods gave results in good agreement with the regular
grid results.

\section{Summary}\label{section:summary}
Using the APEX/LABOCA 870-$\mu$m map of the ECDF-S (Wei\ss~et al.\ 2009) along
with the publically available MUSYC survey data of this field (Taylor et al.\
2008), we have performed a submm stacking analysis of 8266 $K$-band selected
($K_{\mbox{\tiny{vega}}}\le 20$) galaxies, as well as subsets of 737 DRG, 1253 ERO, and
744/149 sBzK/pBzK galaxies. Photometric redshifts have been
derived for the full $K_{\mbox{\tiny{vega}}}\le 20$ sample using $UBVRIzJHK$ data from
MUSYC, thereby allowing us to study stacked submm flux densities as
a function of redshift.  This represents the largest submm
stacking analyses of near-IR selected galaxies to date.  The main results are
summarized below:
\\

$\bullet$ We measure stacked 870-$\mu$m signals of $0.20\pm 0.01\,$mJy
($20.0\sigma$), $0.45\pm 0.04\,$mJy ($11.3\sigma$), $0.42\pm 0.03\,$mJy
($14.0\sigma$), and $0.41\pm 0.04\,$mJy ($10.3\sigma$) for the
$K_{\mbox{\tiny{vega}}}\le 20$, BzK, ERO and DRG samples, respectively.
Splitting the BzK galaxies up into star-forming (sBzK) and passive (pBzK)
galaxies, the former is significantly detected ($0.48\pm 0.04\,$mJy,
$12.0\sigma$) while the latter,as expected, is not ($0.27\pm 0.10\,$mJy,
$2.7\sigma$). This implies that $K_{\mbox{\tiny{vega}}}\le 20$ galaxies are
responsible for $15\pm 5$\% of the EBL at 870-$\mu$m. sBzK galaxies, EROs and
DRGs (brighter than $K_{\mbox{\tiny{vega}}}\le 20$) are found to contribute
$\sim 4\%$ of the 870-$\mu$m background each.
\\

$\bullet$ Performing the stacking analysis in redshift bins, it is found that
the stacked submm signal from $K_{\mbox{\tiny{vega}}}\le 20$ galaxies, as well
as the ERO and DRG sub-samples, is coming from sources in the redshift range
$1.4\ls z \ls 2.5$, while for BzK galaxies the signal remains constant with
redshift.  Assessing the contribution to the submm EBL from the different
samples as a function of redshift is complicated by the fact that these are
flux-limited samples, although, we can conclude that the bulk of the submm
light from $K_{\mbox{\tiny{vega}}}\le 20$ galaxies comes from sources at $z>1$.
\\

$\bullet$ We find a linear correlation between stacked submm flux and 24-$\mu$m
flux density for sBzK, ERO and DRG galaxies with $S_{24\mu m}\ls 350\,\mu$Jy.
This correlation suggests that the 24-$\mu$m emission from $S_{24\mu m}\ls
350\,\mu$Jy galaxies is dominated by star formation, and consequently can be
used as a robust tracer of star formation.  At $S_{24\mu m} > 350\,\mu$Jy we
find that the stacked 870-$\mu$m flux density becomes constant ($\sim
1.3\,mJy$) and independent of $S_{24\mu m}$, which is likely due to AGN
starting to contribute significantly to the 24-$\mu$m as well as 870-$\mu$m
emission.
\\

$\bullet$ In an effort to isolate a subset of BzK, ERO, and DRG galaxies
responsible for the bulk of the stacked 870-$\mu$m emission we have measured
the significance of the stacked submm signal across the BzK and RJK diagrams,
identifying the regions of strongest submm emission. We find that the stacked
submm signal from submm-faint sBzK galaxies is dominated by the subset of sources which
also fulfill the ERO criterion. These are likely to be dusty, starforming galaxies, which are
sufficiently bright in the blue to be selected as sBzK galaxies. The majority of these sources
are found in the redshift range $z=1.5-3$, in line with the BzK-selection criterion proposed by
Daddi et al.\ (2004).
Guided by the stacked submm-contours we propose slightly modified BzK-selection criteria, namely
\begin{eqnarray}
(z-K)_{\mbox{\tiny{AB}}} &\ge& (B-z)_{\mbox{\tiny{AB}}} -0.5\\ 
(B-z)_{\mbox{\tiny{AB}}} &\le& 3.4\\
\end{eqnarray}
for sBzKs, and
\begin{eqnarray}
(z-K)_{\mbox{\tiny{AB}}} &\ge& 2.9\\
(B-z)_{\mbox{\tiny{AB}}} &>& 3.4\\
\end{eqnarray}
for pBzKs. 

In the RJK-diagram we find that the strongest submm signal comes from galaxies in the ERO/DRG overlap
region with $(J-K)_{\mbox{\tiny{AB}}}> 1.9$, which are predominantly found at $z\ge 1.5$.
\\

\acknowledgments
We are grateful to Loretta Dunne and Ryan Quadri for providing us with the
photometric redshift distributions published in D08 and Q07.  IRS acknowledges
support from the Royal Society and STFC. TRG is grateful to the European
Southern Observatory (ESO) for sponsoring a 4 months visit in Garching during
which much of the analysis presented here was carried out, and would in
particular like to thank Sune Toft, Carlos de Breuck and Jesper Sommer-Larsen
for useful discussions and suggestions during this period.

\end{document}